**Effects of Self-Hybridized Exciton-Polaritons on $WS_2$ Photovoltaics**

Adam D. Alfieri, Tobia Ruth, Cheryl Lim, Jason Lynch, Deep Jariwala*

Department of Electrical and Systems Engineering, University of Pennsylvania, Philadelphia, PA 19104, USA

*Corresponding Author: dmj@seas.upenn.edu

## Abstract:

Excitonic semiconductors such as transition metal dichalcogenides (TMDCs) are attractive for next-generation photovoltaics (PVs) with low cost, light weight, and low material consumption. In $WS_2$ and other TMDCs, the simultaneous large optical constants and strong exciton resonance can result in the primary photogenerated species being self-hybridized exciton-polaritons emerging from the strong coupling of excitons and optical cavity modes formed by the $WS_2$. We show that strong coupling can benefit photovoltaic performance, with external quantum efficiencies and power conversion efficiencies enhanced by an order of magnitude, approaching values of 55 and 2%, respectively. Thickness dependent device characterization is performed to study the polariton dispersion, revealing anomalous internal quantum efficiency and fill factor behavior that are attributed to polariton-modified exciton transport processes. Our results uncover a significant mechanism in the photoconversion process for PVs from high index, excitonic semiconductors and indicate the utility of strong coupling for optoelectronic devices.

## Introduction:

Organic semiconductors and various other excitonic nanomaterials have emerged as promising candidates for next-generation ultrathin PVs[1–8]. Current in excitonic PVs is generated by exciton diffusion to heterointerfaces where the excitons dissociate into free carriers[9]. However, exciton diffusion lengths tend to be short (typically <30 nm), so exciton transport is a key limitation on the performance of excitonic PVs. Improving exciton transport is therefore crucial to unlock the potential of excitonic nanomaterials for photovoltaic energy conversion.

Exciton-polaritons are part-light, part-matter hybrid states arising from the strong coupling (SC) of excitons and photons in an optical cavity[10–12]. Due to the cavity photon component, the effective mass of exciton-polaritons is orders of magnitude smaller than the effective mass of bare excitons[13]. Consequently, long-range ballistic polariton propagation at non-zero incident angles has been demonstrated[14,15]. SC has also enabled long-range resonant energy transfer in donor/acceptor systems where both the donor and acceptor excitons are strongly coupled to the cavity photon[16–18]. Early demonstrations of organic semiconductor devices in closed-cavity systems have shown preliminary evidence that SC can enhance quantum efficiency of PVs[19,20], but a clear relationship between SC and internal quantum efficiency has yet to be proven, particularly in the absence of a donor/acceptor system.

Exciton-polaritons are often explored in closed cavity Fabry Perot (FP) resonators with an excitonic medium sandwiched between a nearly perfectly reflecting bottom mirror and a partially transparent top mirror, but films with simultaneously strong exciton resonances and sufficiently high refractive index can form self-hybridized exciton-polaritons in open cavity systems[21–23].

Using a self-hybridized system enables off-resonance absorption in the active layer, improving the practicality of exciton-polariton devices for energy harvesting.

Transition metal dichalcogenides (TMDCs) are van der Waals (vdW) materials that exhibit strong interaction with light due to large exciton oscillator strengths, even in bulk form[24,25], and high refractive indices, enabling strong absorption in thin layers[5,26,27]. Therefore, TMDCs have attracted significant attention as materials for next generation optoelectronics with thin, stable, and non-toxic active layers[28–31]. Self-hybridized exciton-polaritons have been demonstrated in the prototypical disulfides and diselenides of Mo and W, with Rabi splitting values of approximately 200 meV ($4.8 \times 10^{13}$ Hz)[32]. The speed of Rabi oscillations, and therefore the light-matter energy exchange, will occur faster than exciton dissociation and decay processes. Consequently, it is important to understand how self-hybridized exciton-polaritons can affect the photocurrent generation process, both for the design of TMDC PVs and more generally as a model system that can be extended to other high-index excitonic semiconductors, both known and to be discovered.

In this work, we leverage both the large optical constants and excitonic properties of $WS_2$ to fabricate photovoltaics that exhibit self-hybridized exciton-polaritons to understand how they (i) affect TMDC photovoltaics and (ii) can be more broadly applied to other excitonic materials to overcome limitations of exciton diffusion length. We fabricate dozens of devices over a large thickness range to alter the cavity energy and cover the polariton dispersion. We show that exciton-polaritons in self-hybridized systems can enable enhanced efficiencies in PVs, opening new avenues in photovoltaic science and engineering.

## Results and Discussion:

### *Device Structure and Exciton-Polariton Dispersion*

Figure 1a schematically shows the possible photocurrent generation mechanisms in our device. First, excitons created in the bulk can diffuse to an interface at which the excitons will dissociate to free carriers. One carrier is collected and the other undergoes diffusive transport to the opposite electrode. The localized nature of excitons limits the exciton diffusion length and efficiency of this process, particularly out of plane. Second, excitons created in the bulk can ionize to free carriers and undergo drift-diffusion transport. Finally, we consider how strong coupling can modify the diffusive transport of excitons. Here, the excitations are superpositions of photonic modes and excitonic wavefunctions. The effective mass, and therefore the transport, is dominated by the photonic component[13], while the charge is effectively encoded in the exciton. Because the energy oscillates between the photon component and the exciton component at the Rabi frequency, we can consider that polaritonic transport occurs through one (or multiple) Rabi oscillations in which excitons undergo one half of a Rabi oscillation and exchange energy to the photon; undergo photon-like transport; undergo another half Rabi oscillation back to the exciton; and dissociate at a charge-selective contact. One charge is immediately extracted while the opposite charge undergoes drift-diffusion transport to the other contact.

We fabricate PVs using mechanically exfoliated $WS_2$ flakes on 2 nm Pt/70 nm Au/5 nm Ti bottom electrodes with a $WO_x$/CVD graphene electron selective contact[28] (Figure 1b, see Methods for details). 10 nm Ti/200 nm Au electrodes "patch" the graphene to the top electrode contacts. Figure 1c shows a representative device. Photocurrent mapping (Figure S4.) confirms that the device active area is the region where the $WS_2$ is both directly on top of the bottom electrode and is covered by graphene.

Figure 1d shows the experimentally measured and transfer matrix method[33] (TMM)-calculated reflectance spectra for devices with various thicknesses of $WS_2$, showing the accuracy of TMM. Figure 1e shows the TMM-calculated reflectance as a function of photon energy and $WS_2$ thickness to study the exciton-polariton dispersion. Minima in measured reflectance spectra for representative devices are overlaid as orange dots, showing excellent agreement between experiment and TMM calculation. The polariton modes are easily identified as the sharp reflectance minima that feature an anti-crossing between the cavity mode and exciton energy. The polariton modes are fit as the eigenvalues of the multimode, multiexciton coupled oscillator Hamiltonian for a system with $l$ FP cavity modes and $m$ exciton states (Methods)[34,35]. The energies of the FP cavity modes generally depend on thickness and incident angle. However, the massive optical impedance of $WS_2$ results in negligible dispersion with incident angle (Figure S3.). Therefore, all measurements are conducted at normal incidence, and the cavity tuning is modified only by changing the $WS_2$ thickness.

In addition to the $\sim l\lambda/(2n)$ (with $l$ an integer and $n$ the refractive index) FP cavity modes, there is a deep subwavelength interference mode[36] for thinner (<20 nm) films that interacts with the exciton in the weak coupling regime, leading to a Fano resonance that causes the blue shifting of the reflectance minima from the bare exciton peaks for thicknesses below 50 nm[37]. We focus on the $\sim l\lambda/(2n)$ cavity modes where $l$ is 1, 2, and 3, which all exhibit SC (Supporting Information), and we further limit our focus to the $WS_2$ A exciton (approximately 630 nm) as it has the largest oscillator strength, narrowest linewidth, and lowest energy. For the remainder of this work, we refer to the polariton branch between 525-630 nm as the upper polariton (UP) and the polariton branch beyond 630 nm as the lower polariton (LP). The Rabi splitting for the first 3 polariton modes are 195.8 meV, 196.9 meV, and 196.0 meV, respectively.

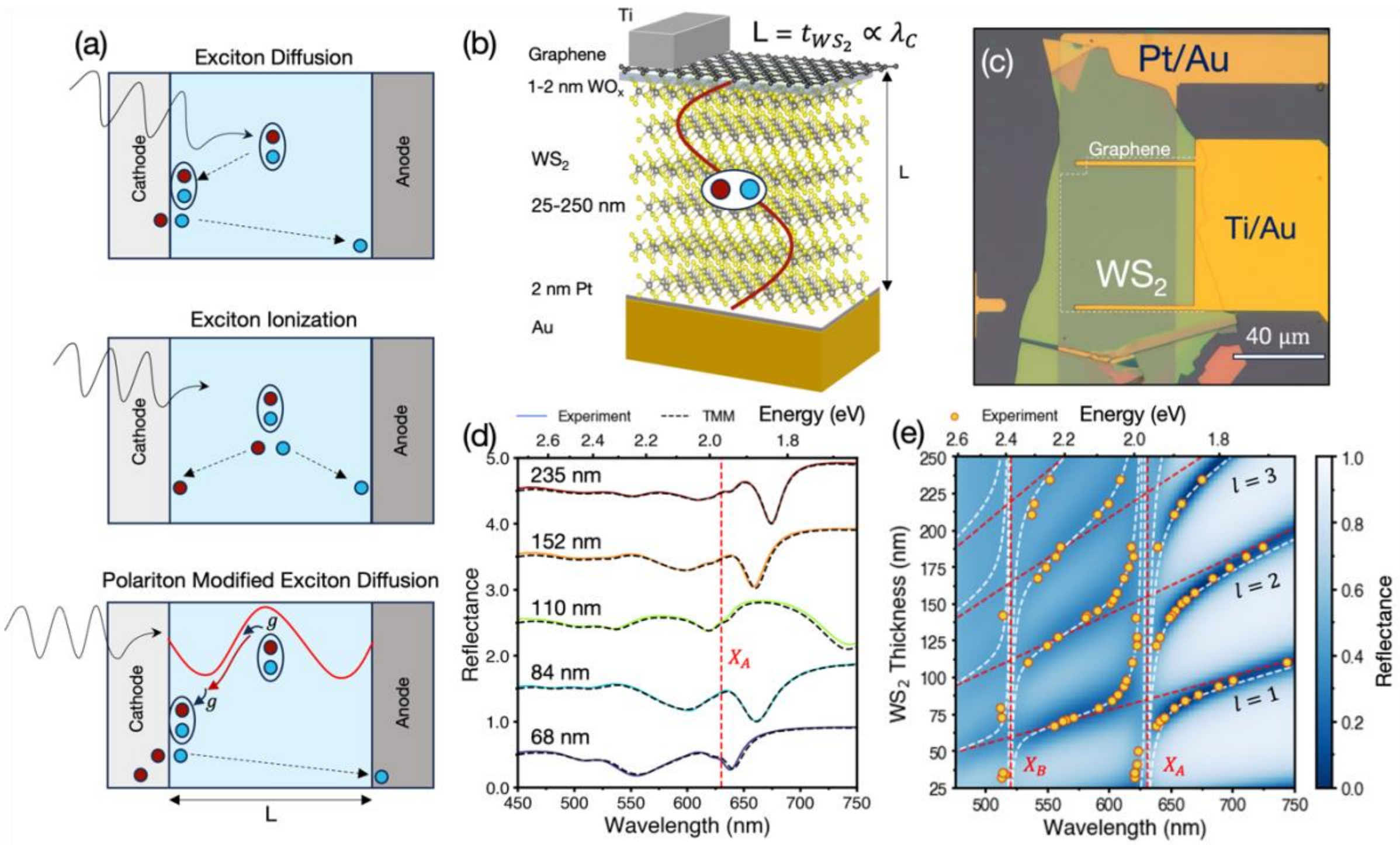

***Figure 1. Exciton polariton photovoltaics, device structure, and polariton dispersion.*** (a) Schematic comparison of photocurrent generation processes based on exciton diffusion transport, exciton ionization, and polariton modified exciton diffusion. (b) The structure of the exciton-polariton PV devices. Bulk $WS_2$ forms an optical cavity when placed on a 2 nm Pt/70 nm Au bottom electrode. A thin $WO_x$/graphene electron selective bilayer with 10 nm Ti/200 nm Au finger electrodes forms the top contact. (c) An optical microscope image of a representative device. (d) Measured reflectance spectra (solid lines) and TMM calculated spectra (dashed lines) for various thicknesses. Spectra are shifted vertically for clarity. The red dashed line corresponds to the A exciton. (e) Reflectance calculated by TMM as a function of thickness and wavelength. Fits to the coupled oscillator model (Eq. 1) are overlaid as dashed white lines, while the red lines are the bare cavity photons and the exciton energies. The cavity mode orders are labeled, and the A and B excitons are denoted. The minima in the reflectance spectra for samples of various thicknesses are overlaid as orange dots.

### *Spectral Efficiency*

We now investigate the effects of polaritons on external quantum efficiency (EQE) and power conversion efficiency (PCE) as a function of wavelength. Figures 2a and 2b shows the (EQE) and PCE spectra for representative devices near zero detuning, $\Delta_l(t) = E_C^{(l)}(t) - X_A$ (where $E_C^{(l)}(t)$ is the energy of cavity mode $l$ and $X_A$ is the energy of the A exciton), for the first 3 FP cavity mode polaritons (see Figure S6. for quantitative EQE and PCE values). Each EQE and PCE spectrum exhibits clear peaks corresponding to the UP and LP resonances. This is clearly shown in Figures 2c and 2d, which plot the energies of the peaks for each EQE (2c) and PCE (2d) spectra as a function of thickness. The peaks are overlaid over the polariton dispersion and follow the dispersion of the polariton modes remarkably well, evidencing the beneficial effect of polariton states on photovoltaic performance. Multiple peaks can emerge in the spectra due to multiple polariton mode orders. The clear resemblance of PCE extracted from monochromatic I-V characteristics with polaritonic dispersion presents strong evidence of the effect of exciton-polaritons directly on the PCE of PVs.

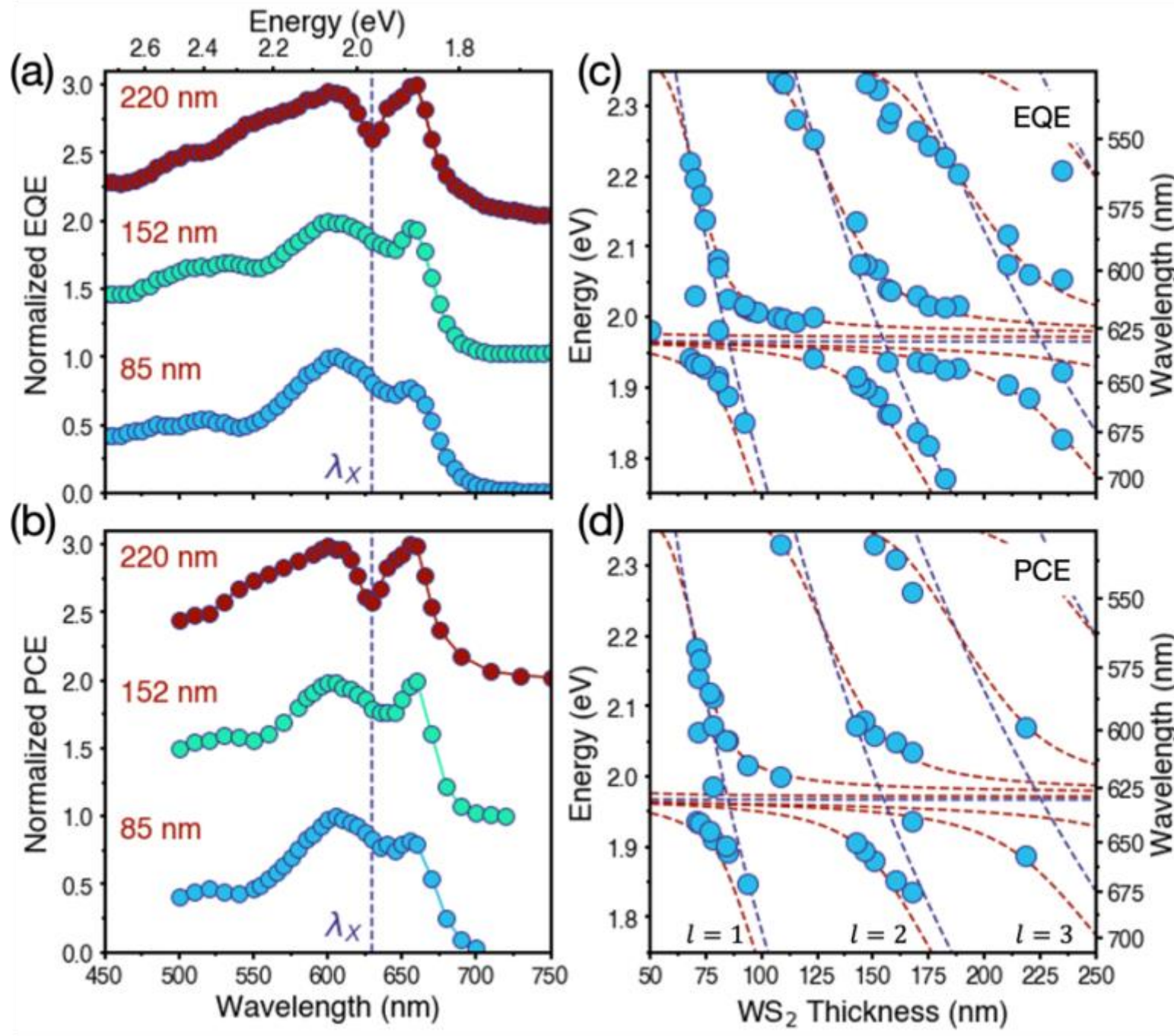

***Figure 2. Spectral EQE and PCE***. Self-normalized EQE (a) monochromatic PCE (b) spectra for representative devices with $WS_2$ thicknesses of 85 nm, 152 nm, and 220 nm, corresponding to approximate zero-detuning thicknesses for polaritons arising from the first, second, and third order FP cavity modes, respectively. The spectra are shifted vertically for clarity. The wavelength corresponding to the bare exciton, $\lambda_X$, is denoted by the dark blue dashed line. Peaks in the EQE (c) and PCE (d) spectra as a function of $WS_2$ thickness overlaid on the polariton dispersions (red dashed lines) and the bare exciton and cavity energies (blue dashed lines).

### *Internal Quantum Efficiency*

With a Rabi splitting of 196 meV (4.74 x $10^{13}$ Hz), strong coupling will proceed faster than the theoretically predicted ionization rate for excitons in $WS_2$ under expected operating conditions (<10 V/μm) by at least an order of magnitude[38]. Therefore, exciton-polaritons are certainly expected to affect the charge conversion process in addition to improving absorption. To study this phenomenon, a thickness dependent study of internal quantum efficiency (a measure of transport and charge extraction) is needed. Figure 3 shows the thickness dependent EQE (top) and internal quantum efficiency (IQE, bottom) of devices at wavelengths of 450 nm (off resonance), 600 nm (near UP when $\Delta = 0$), 630 nm (near bare exciton), and 650 nm (near LP when $\Delta = 0$). The LP wavelength at $\Delta_l(t_l) = 0$ is approximately 660 nm for all $l$, but we are limited by the bandwidth of our tunable laser filter, hence the choice of 650 nm. Vertical dashed lines are included to indicate the thicknesses corresponding to $\Delta_l(t_l) = 0$ for the $l$ = 1,2, and 3 FP cavity modes, respectively. We denote the weak coupling (WC) regime with red shading. The cutoff is chosen to be at a point where the first cavity mode energy is detuned $4g$ above the exciton energy, resulting in an exciton fraction of approximately 95% for the LP. To eliminate uncertainty related to the active area, we use a focused, tunable laser beam with a spot size of approximately 2 μm in diameter (see Methods) at 25 points on each device and extract the average and standard deviation of the short circuit current. The devices are then grouped by thickness and the group mean of the EQE/IQE are plotted in Figure 3 with the error bars as the standard error. Figure S8 shows another version of Figure 3 in which the mean and standard deviation of each individual device are plotted.

Due to the electric field enhancement of the polariton states, the Au and Pt layers exhibit large parasitic absorption peaks (Figure S2.). To understand the effect of the polaritons on transport, we must only consider the absorption in the $WS_2$ and graphene active layers. The absorption in the active layer must be extracted from layer-resolved absorption with the transfer matrix model (Figure S2.). The IQE is simply the EQE normalized by the active layer absorption and therefore a measure of excited state charge-transport efficiency inside the PV.

$$IQE(\lambda) = \frac{EQE(\lambda)}{A_{active}(\lambda)}$$

The EQE and IQE both increase once the thickness enters the SC regime. The increase in the EQE from the WC regime to the SC regime is by a factor of as much as 10 near the polariton resonances, ~5 at the exciton wavelength, and ~3 for off-resonance illumination. The increase and oscillatory behavior of the EQE at 600 nm and 650 nm are expected based on our results in Figure 2, further solidifying the benefit of SC on EQE. The IQE in the SC regime is ~3 times that of the WC regime. At all 4 wavelengths, there is a prominent peak in the IQE near $t_l$, with IQE approaching unity for both 630 nm and 650 nm illumination. IQE values > 90 % in our devices (Figure S8) with thickness corresponding to $l$ =1 are among the highest experimentally measured values reported for TMDC PVs[3,5,30,39].

Bulk recombination, surface recombination, and generation profiles are normally the primary parameters that would affect the thickness dependence of the IQE. The active layers here are prepared by top-down mechanical exfoliation from the same parent single crystal, and the thicknesses considered here are all well beyond the thicknesses at which quantum confinement influences the electronic structure of $WS_2$ (~5 nm and below)[40]. Therefore, the bulk free carrier mobilities and lifetimes are thickness independent. The surface recombination velocities will also be thickness independent due to the identical device fabrication process. We perform rigorous drift diffusion (DD) modelling to identify potential origins of the thickness dependent IQE. However, free-carrier and conventional exciton modelling with a constant diffusion length are unable to replicate the observed behavior (Figures S10-S15). However, by fitting the exciton diffusion length to the EQE in the WC and SC regimes individually (dark blue and light blue dashed lines in Figure 3), we can see that drastically improved exciton diffusion in the SC regime relative to the WC regime begins to better represent the experimental results. Therefore, we expect that SC can have a beneficial effect on transport and conclude that the photocurrent generation process here is best described by a thickness dependent exciton diffusion length that arises from SC. In addition to this effect being observed at excitations on/near resonance, we observe the SC enhanced exciton diffusion for the off-resonance (450 nm) excitation, though the efficiencies here are low, possibly the result of inefficient exciton generation. The presence of the thickness dependent transport suggests a process of relaxation to polariton states: a phenomenon that is well known from studies of polaritonic light emission[22,23,41]. This is a crucial finding that further underscores the importance of self-hybridized devices as it means that broadband absorption above the bandgap can adopt the benefits of polaritonic transport.

We develop a model for thickness dependent exciton diffusion (TDXD) in the devices mediated by polaritons. First, a diffusion model combines the concepts of resonance energy transfer (RET)[16], homogeneous RET[42,43], and diffusive transport by hopping[44] to obtain a phenomenological diffusion coefficient that describes a mechanism by which SC could alter exciton diffusion. The model includes the phenomenological RET diffusivity at polariton branches weighted by the thickness-dependent exciton fraction for each polariton branch. The thickness dependent diffusion length is then input to the exciton diffusion current model to get the calculated EQE, which is fit to the experimental EQE. A detailed discussion of the theory is included the Supporting Information. The model predicts the shape of the EQE/IQE better than free carrier or traditional exciton models, particularly the transition from WC to SC, despite the complexity of the system and the relative simplicity of the model. The relative accuracy of the TDXD model adds to evidence that polaritons modify exciton transport. Realistically, the system will have both polaritons and free carriers, which could explain deviations of our theory at larger thicknesses, but our results suggest that SC has a significant effect on the quantum efficiency of $WS_2$ PVs and must be considered.

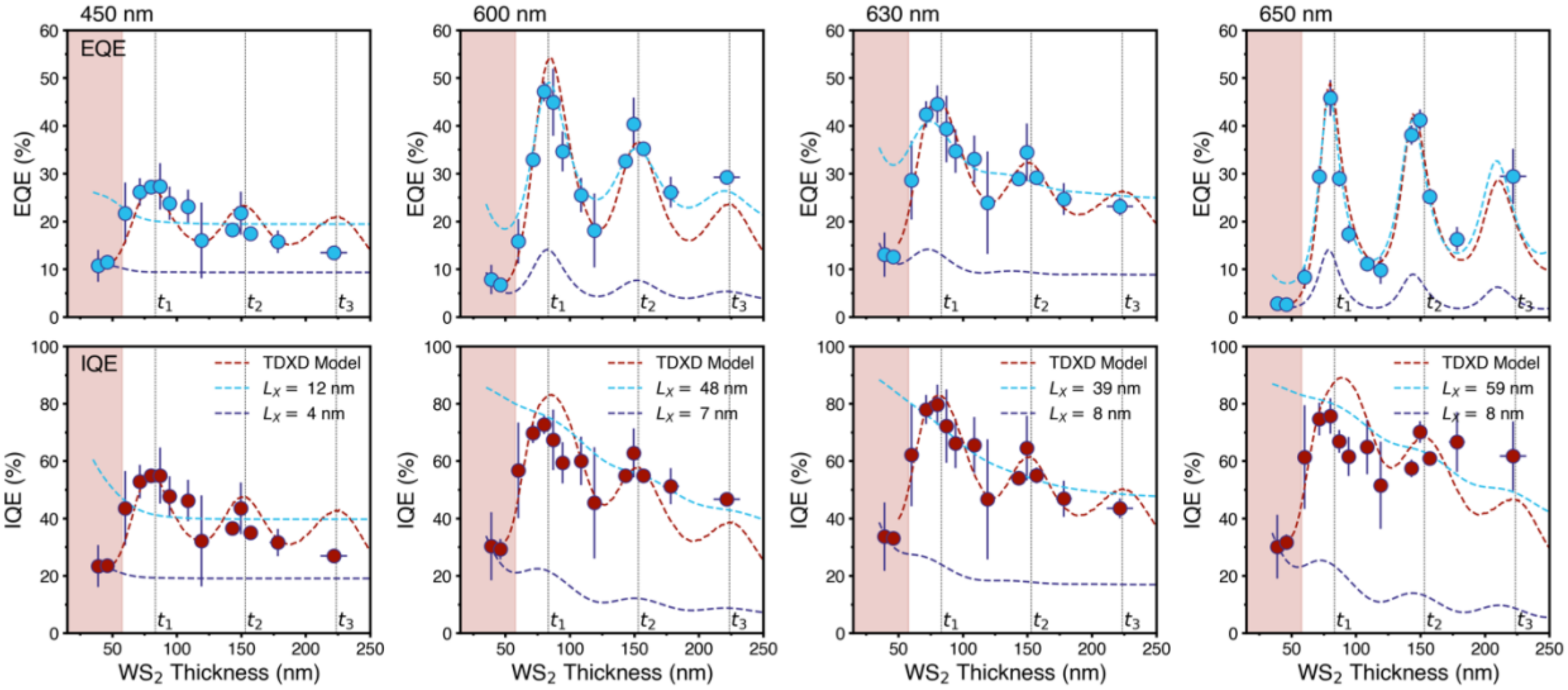

***Figure 3. Thickness dependent EQE and IQE.*** External (top) and internal (bottom) quantum efficiencies determined by photocurrent mapping with a tunable laser at wavelengths of 450 nm (a), 600 nm (b), 630 nm (c), and 650 nm (d). The weak coupling (WC) regime is denoted by red shading while the strong coupling (SC) regime is unshaded. The EQE and IQE predicted by the thickness dependent exciton diffusion length (TDXD) model are included as the red dashed lines. Fits to the exciton diffusion current model with constant diffusion lengths for the WC and SC regime are included as the dark blue and light blue dashed lines. The thicknesses, $t_l$ ($l$ = 1,2,3), corresponding to zero detuning of the $l$=1,2,3 cavity modes are denoted with black dotted lines. Each data point represents the mean and standard error of a group of devices. The mean and standard deviation of the individual devices are provided in Figure S8.

### *Photovoltaic Characteristics*

Finally, we consider photovoltaic measurements under solar simulation conditions (see Methods). Figure 4a shows light and dark J-V curves for the champion device, which has a PCE of 2.0%, comparable to other TMDC photovoltaics and the highest performing $WS_2$ PV to date. In Figure 4b, dark J-V curves of representative devices of various thickness are plotted on a semi-log plot. The lower slope of the semi-log J-V curve for some thicker devices at lower and higher voltages indicates increased free carrier recombination in the bulk and greater series resistance. Otherwise, there are not significant differences with thickness apart from random device-to-device variation.

In Figures 4c-f, short circuit current density ($J_{sc}$), open circuit voltage ($V_{oc}$), fill factor (FF), and PCE are plotted as a function of $WS_2$ thickness. Devices are grouped by thickness to account for device-to-device variation, and the mean and standard deviation are plotted. $J_{sc}$ is most strongly related to thickness and SC due to increased absorption and the aforementioned benefits of polaritonic transport. This effect is confirmed by maxima at the thicknesses of $t_l$ and the strong initial increase in $J_{sc}$ as the SC regime is entered, which exceeds the amount of increase predicted by DD modelling. $V_{oc}$ shows no clear relationship with polaritons (Figure 4d and Figure S10.). Our DD modelling suggests the large $V_{oc}$ deficit is likely dominated by non-ideal contacts and Fermi-level pinning rather than recombination (Figure S12), though it does suggest that a slight drop in $V_{oc}$ could be expected with decreasing thickness. FF (Figure 4e) reflects the recombination in the device. An increase with thickness is observed, with a FF that increases in the SC regime. Like our IQE results, this trend is not predicted by conventional PV modelling, which generally predicts that FF will decrease monotonically with thickness, even in cases of large surface recombination (Figure S10). Taken with our IQE results, the FF vs thickness trend further supports our claim that SC can reduce recombination in PV devices through improved transport.

The PCE (Figure 4f) is directly proportional to $J_{sc}$, $V_{oc}$, and FF:

$$PCE = \frac{J_{sc} V_{oc} FF}{P_{inc}}$$

Therefore, we expect polaritons to enhance PCE given the increase in $J_{sc}$ and FF and no detrimental impact on $V_{oc}$. An increase of PCE in the SC regime reflects the expected behavior. The PCE values in the SC regime are at least an order of magnitude larger than the best device in the WC regime, which has a PCE of approximately 0.13%. The magnitude by which the PCE is enhanced exceeds what is predicted by DD modelling, which only reflects the increased absorption in the SC regime and a modest decrease in $V_{oc}$ at lower thicknesses due to surface recombination. PCE values in the 1-2% range are comparable with those reported for TMDC PVs in the past[28–31], but we emphasize the importance of the thickness dependent trend and not the absolute magnitudes. Design of future TMDC PVs must account for strong coupling and should have a thickness near the respective $t_1$ or $t_2$ for maximum efficiency.

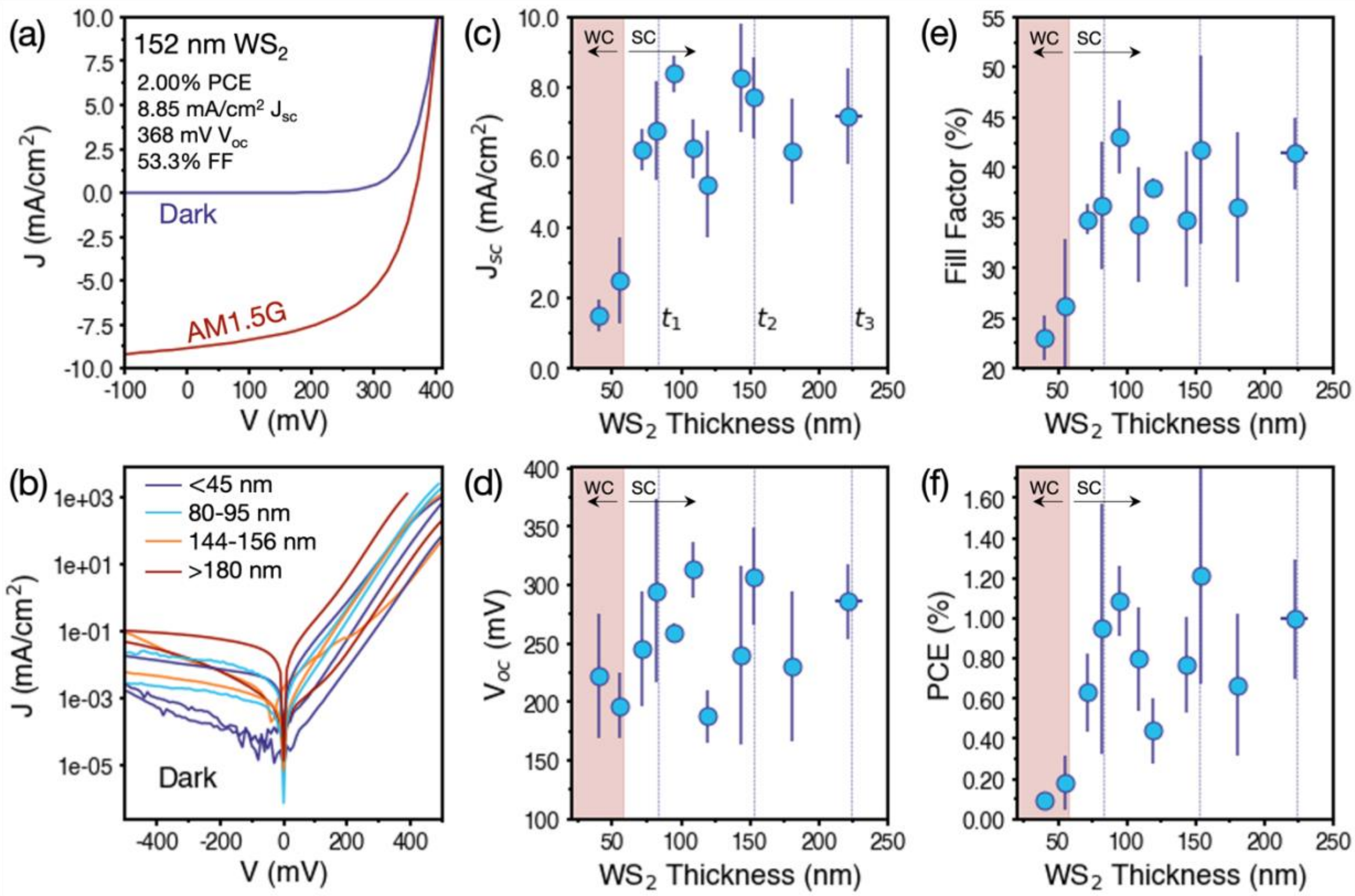

***Figure 4. Thickness dependent photovoltaic characteristics.*** (a) Light and dark J-V curves for a champion device. (b) Semi-log dark J-V curves for representative devices in various thickness ranges. Short circuit current (c), open circuit voltage (d), fill factor (e), and PCE (f) as a function of thickness. The total incident power is 86.7 mW/cm$^2$. The thicknesses, $t_l$ ($l$ = 1,2,3), corresponding to zero detuning of the $l$=1,2,3 cavity modes are denoted with black dotted lines. In figures (c-f) devices are grouped by thickness and the average and standard deviation are plotted. Individual device data are provided in Figure S17.

## *Outlook and Conclusions*

In conclusion, our work presents self-hybridized exciton-polariton photovoltaics using bulk $WS_2$ as both the cavity and active excitonic material: the first demonstration of self-hybridized exciton-polariton optoelectronic devices. We demonstrate that exciton-polaritons can have a beneficial effect on both the external quantum efficiency and power conversion efficiency of excitonic photovoltaics. Further, we show that the benefit of polaritons is due to not only improved absorption, but also improved internal quantum efficiency and fill factor for devices in the strong coupling regime. These results indicate that self-hybridization should be accounted for in the design and modelling of TMDC PVs. Finally, the possible simultaneous presence of broadband absorption and strong coupling enhanced exciton transport make self-hybridized exciton-polariton devices attractive for solar photovoltaics and other forms of optical energy transduction.

**Acknowledgements**

A.A. and T.R. recognize primary support from the Vagelos Institute for Energy Science and Technology. A.A. and D.J. acknowledge partial support from the Office of Naval Research (ONR) Young Investigator Award (YIP) (N00014-23-1-203) Metamaterials Program. D.J. and J.L. recognize primary support Asian Office of Aerospace Research and Development of the Air Force Office of Scientific Research (AFOSR) FA2386-20-1-4074. D.J. acknowledges support from the Alfred P. Sloan Foundation's Sloan Fellowship in Chemistry. C.L. acknowledges support from the University of Pennsylvania Center for Undergraduate Research Fellowships. This work was conducted in its majority at the Singh Center for Nanotechnology at the University of Pennsylvania, which is supported by the NSF National Nanotechnology Coordinated Infrastructure Program grant no. NNCI1542153. The authors gratefully acknowledge the use of facilities and instrumentation (G2V Pico solar simulator) supported by the Department of Materials Science and Engineering Departmental Laboratory at the University of Pennsylvania. A.A. thanks Zirun Han and Mahfujur Rahaman for helping with aspects of experimental setup.

**Author contributions:**
D.J. conceived, supervised, and acquired funding for the project. A.A. and D.J. designed the experiments. A.A. fabricated the devices with assistance from T.R. and C.L. A.A., T.R., and C.L. performed the optoelectronic measurements. A.A. conducted the ellipsometry measurements and J.L. fit the ellipsometry data. A.A. performed the theoretical calculations, simulations, data processing, and data visualization. A.A. wrote the manuscript with inputs from all authors. All authors discussed the results and revised the manuscript.

**Competing interests**
The authors declare no competing interests.

**Data and materials availability:**
Data, materials, and code are available from the corresponding author upon reasonable request.

# Supporting Information for

## Effects of Self-Hybridized Exciton-Polaritons on $WS_2$ Photovoltaics

Adam D. Alfieri, Tobia Ruth, Cheryl Lim, Jason Lynch, Deep Jariwala

Corresponding author: dmj@seas.upenn.edu

**The PDF file includes:**

**Methods**

*Device fabrication*

Electrodes are patterned on 300 nm thermal $SiO_2$/Si substrates using photolithography, electron-beam evaporation of 5/70 nm Ti/Au (Kurt J. Lesker), then DC sputtering (Kurt J. Lesker) of ~2 nm Pt before liftoff. $WS_2$ flakes are then mechanically exfoliated and transferred to the bottom electrode before annealing in a tube furnace at 75° C for 2 hrs under Ar+15% $H_2$ gas to improve adhesion and remove PDMS residues. The samples are then cleaned with acetone, isopropyl alcohol (IPA), and an $N_2$ dry. An ultrathin (likely < 2 nm) native $WO_x$ layer is formed by treating the samples with an $O_2$ plasma (25W, 100 sccm $O_2$ flow) for 15 seconds (March Jupiter II RIE). PMMA-coated CVD graphene (Graphenea) is wet-transferred from Cu foil to the samples, covering the $WS_2$ flakes. The PMMA is removed in acetone and IPA with an $N_2$ dry. The top "patch" electrodes are then prepared by E-beam lithography (Elionix ELS-75), E-beam evaporation of 10/200 nm Ti/Au and liftoff in acetone. Finally, the graphene is patterned using EBL and an $O_2$ plasma (100W, 100 sccm $O_2$ flow) for 60s before removal of remaining PMMA resist with acetone/IPA/$N_2$.

*Reflectance spectroscopy*

Normal incidence reflectance spectroscopy is performed with a 50x objective (Olympus SLMPN 50X N.A. = 0.35) in ambient conditions using a Horiba LabRam HR Evolution confocal microscope and an external white light source (AvaLight-HAL). The spot size is approximately 4 µm in diameter. An alumina-coated silver mirror serves as the baseline.

*Thickness estimation and Optical Modelling*

Given the accuracy of the TMM calculations, we estimate thicknesses by fitting the thickness-dependent calculated reflectance spectra to the experimentally measured reflectance spectra by a least-squares optimization method. The optical model used is 1 nm graphene/$WS_2$/1.8 nm Pt/70 nm Au/300 nm $SiO_2$/Si. The refractive indices of $WS_2$ and graphene are taken from literature[25,44]. The refractive indices of the Pt/Au bottom electrode are measured via spectroscopic ellipsometry (Figure S1).

*Coupled oscillator model*

The multimode, multi

$$H = \sum_{l} \left\{ \left( E_C^{(l)} + i\gamma_C^{(l)} \right) |l\rangle\langle l| + \sum_{m} \left[ \left( E_X^{(m)} + i\gamma_X^{(m)} \right) |m\rangle\langle m| + g_{ml} \left( |l\rangle\langle m| + |m\rangle\langle l| \right) \right] \right\}$$

Here, $E_C^{(l)}$ and $2\gamma_C^{(l)}$ are the energy and full width half maximum (FWHM) of the $l^{\text{th}}$ order FP cavity mode, which depends on the cavity thickness, $t$; $E_X^{(m)}$ and $2\gamma_X^{(m)}$ are the exciton energy and FWHM; and $g_{lm}$ is the coupling strength of the $m^{th}$ exciton to the $l^{\text{th}}$ order cavity mode[13]. The cavity wavelength is expected to be approximately linear with thickness, so $E_C^{(l)}$ becomes:

$$E_C^{(l)} = \frac{hc}{\lambda_C^{(l)}} = \frac{hc}{(\text{Slope})_l * t + (\text{Intercept})_l}$$

We consider the interference ($l$ = 0) and first 4 FP cavity modes, and neglect interactions between modes. In addition to the A exciton, we also include the B and C excitons of $WS_2$ to account for dispersion at higher photon energies, although the contribution from these states is weak due to the large continuum of conduction band states.

The decay rate of the bare exciton, $\gamma_X$, is estimated by fitting a Lorentz oscillator to the exciton peak in the imaginary permittivity. The full width half maximum (FWHM) of the oscillator is $2\gamma_X$. We estimate $\gamma_X$ to be approximately 25 meV. For each cavity photon mode, a Lorentz oscillator is fit to the bare cavity resonance at a strong negative detuning (when the cavity wavelength (energy) is 800 nm (1.54 eV)).

The polariton branches we fit to are identified as thickness dependent dips in the TMM-calculated reflectance and sorted to specific modes and branches based on thickness and spectral position. The fit is obtained by a least squares minimization method. The fit parameters are listed in Table S1. The Rabi splitting for the $l^{th}$ order polariton is:

$$\hbar\Omega_{Rabi} = \sqrt{4g_{lm}^2 - \left(\gamma_C^{(l)} - \gamma_X^{(m)}\right)^2}$$

The system is in the strong coupling regime when $g_{lm}^2 > \left(\gamma_C^{(l)} - \gamma_X^{(m)}\right)^2/4$. This case is satisfied for the interactions of the $l \neq 0$ cavity modes with the A and B exciton.

*Spectral EQE and PCE measurements*

Devices are wire bonded to a custom-designed PCB (JLCPCB) for optoelectronic measurements. EQE and spectral PCE measurements are performed using a collimated Xe arc lamp light source (Sciencetech) operating at 200W. The light is focused into a monochromator (Newport) configured with a spectral resolution of 5 nm. The output of the monochromator is collimated to a diameter of approximately 1/2" and is incident on a vertically mounted sample (Figure S5.). A 600 nm long pass filter (Edmund Optics) is used at the output of the monochromator for wavelengths of 610 nm and longer to filter out higher order diffractions. The optical power density is measured using a calibrated Si photodiode power meter (Newport), from which the incident photon flux, $\Phi_{ph}$ (in photons $cm^{-2}$ $s^{-1}$), is extracted:

$$\Phi_{ph}(\lambda) = \frac{\lambda}{hc}\frac{P_{inc}(\lambda)}{A_{beam}} = \frac{\lambda}{hc}P_0(\lambda)$$

The short circuit current, $I_{sc}$ is measured using a Keithley 2450 sourcemeter. The EQE is:

$$EQE(\lambda) = \frac{electrons/s}{incident\ photons/s} = \frac{I_{sc}(\lambda)}{qA_{device}\Phi_{ph}(\lambda)}$$

The experimental setup for spectral PCE measurements is identical to the setup for the EQE measurements, except that at each wavelength, a full I-V sweep is taken from -1V to the open circuit voltage. The voltage step size is 0.1V between -1V to 0V, and then 0.05V from 0V until the current changes sign. The spectral PCE is:

$$PCE(\lambda) = \frac{\max(I(\lambda) * V)}{A_{device} P_0(\lambda)}$$

*Local EQE measurements*

EQE measurements with an illumination spot size of 2 μm are conducted using a supercontinuum laser (NKT Photonics) configured with a RF-driven acousto-optic tunable filter, covering a range of 400 nm to 650 nm. The tunable laser is fiber-coupled to the Horiba confocal and focused with the same 50x objective used for reflectance measurements. Photocurrent measurements are taken by performing photocurrent mapping over 25 spots. The short circuit current is measured by a Keithley 2450 sourcemeter, and the mean and standard deviation are reported. The incident optical power for each wavelength is measured using a power meter (Newport). The experimental setup is shown schematically in Figure S7.

*Solar simulator measurements*

Solar simulation I-V measurements were performed under ambient conditions using a G2V Pico LED solar simulator and a Keithley 2400 sourcemeter. I-V measurements were performed in the dark and under ~1 sun (86.7 mW/cm$^2$) AM1.5G illumination. A control script first automatically finds the open circuit voltage, then the voltage is swept from -1.1$V_{oc}$ to +1.1$V_{oc}$ at 50 points at a scan rate of approximately 0.2 V/s. No preconditioning is performed, but the PCE measurements are performed after the EQE measurements for each device. The active area is estimated to be the region of overlap between the graphene top electrode and Pt/Au bottom electrode, as there is negligible photocurrent outside this region (Figure S4.).

*Photocurrent mapping*

Photocurrent mapping measurements are performed using a Horiba LabRam confocal microscope with a 633 nm He-Ne laser filtered with a 0.1% ND filter. The measurement is conducted at normal incidence with a 50x objective lens (Olympus SLMPN 50X N.A. = 0.35). Short circuit current is measured over an x-y grid at step sizes of approximately 3 μm using a Keithley 2450 sourcemeter. Raman measurements are collected simultaneously. Photocurrent maps and associated Raman maps of the $WS_2$ $A_{1g}$ peak for two devices are shown in Figure S4. The active area is well defined by the region of overlap between the graphene top electrode and the metal bottom electrode.

*Device Simulation*

We perform drift-diffusion (DD) simulation of our system to better understand the effects of polaritons on device characteristics. First, only free-carrier effects are considered, and the generation rate is scaled by a photon efficiency: $\boldsymbol{G_{eff}} = \boldsymbol{\eta_{photon}} \cdot \boldsymbol{G_{abs}}$. This approximation is valid if we neglect the effects of strong coupling on the current generation process, as the dissociation rate of uncoupled excitons is expected to occur faster than excitons recombine – though we reiterate that this dissociation will proceed at a significantly slower rate than Rabi oscillations. This approximation also assumes that the exciton transport prior to recombination is negligible. In our simulations, we let $\boldsymbol{\eta_{photon}} = \mathbf{1}$ as modifying it simply scaled the IQE and EQE linearly.

We consider the effects of surface recombination velocity, carrier mobility, bulk carrier lifetime (assumed to be limited by Shockley-Read-Hall recombination), and contact work functions on the thickness dependence of EQE and IQE (under illumination wavelengths of 450,600,630 and 650 nm) and photovoltaic parameters ($J_{sc}$, $V_{oc}$, FF, and PCE, all under AM1.5G illumination). These parameter sweeps are shown in Figures S10-S13. The range of carrier mobilities is assumed to be between 0.01 $cm^2$/Vs to 1 $cm^2$/Vs due to the flat bands for out of plane transport in TMDCs[1], and the carrier lifetimes are expected to be on the order of 5-100 ns. DD calculation is performed using Sesame[2], an open source solar cell modelling solver from NIST.

Second, we consider purely excitonic photocurrent from diffusion and exciton separation at the front (x = 0) and back (x = d) interfaces. The exciton density profile, $\boldsymbol{X(x)}$, at steady state is given by solving the following differential equation:

$$\frac{d^2X}{dx^2} = \beta^2 X - \frac{1}{D}G(x), X(0) = X(d) = 0$$

Here, $\boldsymbol{\beta}$ is the reciprocal of the exciton diffusion length, $\boldsymbol{L_X} = \sqrt{\boldsymbol{D\tau_X}}$. The exciton density profile is then calculated as given by Pettersen *et al.*[3] as a function of thickness and $\boldsymbol{L_X}$ for illumination wavelengths of 450 nm, 600 nm, 630 nm, and 650 nm. The short circuit photocurrent resulting from exciton diffusion and interfacial dissociation is proportional to the gradient of the exciton density at the front and back interfaces:

$$J_X = qD\left(\theta_0 \frac{dX}{dx}\bigg|_{x=0} - \theta_d \frac{dX}{dx}\bigg|_{x=d}\right)$$

Here, $\boldsymbol{q}$ is the fundamental charge, and $\boldsymbol{\theta_{0,d}}$ are the exciton dissociation/collection efficiencies at the front and back interfaces. The EQE and IQE at illumination wavelengths of 450 nm, 600 nm, 630 nm, and 650 nm are plotted as a function of thickness for various $\boldsymbol{L_X}$ values in Figure S14. These results show that exciton generation profiles alone cannot describe our results, particularly for the 450 nm and 630 nm wavelengths. Figure S15 shows the predicted spectral EQE for different $WS_2$ thicknesses as a function of $\boldsymbol{L_X}$ and $\boldsymbol{\theta_{0,d}}$. The model overestimates the EQE at shorter wavelengths relative to Figure 2 of the main manuscript.

***Theoretical Model for Polaritonic Transport***

Bulk TMDCs have strong confinement of excitons in-plane[4] and low mobilities (~$10^{-2}$ $cm^2$/V-s) for out-of-plane carrier transport[1] due to approximately flat bands along the $K - H$ ($k_z$) direction in the Brillouin zone. Estimates of the exciton effective mass in the out of plane direction are scarcely reported, but one estimate results in a reduced effective mass of ~$2m_0$[5]. One can therefore expect out-of-plane transport to be governed by a hopping mechanism. As in organic small molecule semiconductors, we can expect the hopping to be dominated by Förster resonance energy transfer (FRET) and Dexter energy transfer (DET) mechanisms[6]. The diffusion coefficient for hopping transport is proportional to sums of the products of the energy transfer (ET) length scale and ET rate constant for each mechanism, j:

$$D \propto \sum_{j} L_{ET,j}\Gamma_{ET,j}$$

The FRET rate constant is proportional to the photoluminescence efficiency[6], which is small for bulk $WS_2$ due to the indirect bandgap. We can therefore expect that out-of-plane bare exciton transport is dominated by a DET process between individual layers, which inherently occurs on sub-nanometer length scales, limiting the diffusion coefficient.

We propose a possible mechanism of exciton-polariton transport in which there is polaritonic homogeneous resonance energy transfer (RET) between identical but spatially separated molecules, through which we expect a drastic enhancement in both the rate and length scale for hopping transport. Homogeneous resonance energy transfer processes have proven to enhance exciton diffusion in molecular photonic wires[7] and arrays of aligned colloidal quantum well nanoplatelets[8]. In quantum electrodynamics (QED), RET proceeds through intermediate "virtual photons"[9]. In the case of polaritons, the intermediate photonic states are now the cavity photon modes[10]. This concept is consistent with the very definition of exciton-polaritons, in which exciton-photon dipole-dipole coupling can be thought of as energy exchange through a rapid cycle of absorption and re-emission of cavity photons[11].

*Generalized Resonance Energy Transfer Framework*

In the QED treatment of RET, the system Hamiltonian is:

$$H = H_D + H_A + H_{rad} + H_{int,A} + H_{int,D}$$

$$H_X(X = A, D) = \hbar\omega_X \hat{b}_X^\dagger(z_X)\hat{b}_X(z_X)$$

$$H_{rad} = \hbar\omega_C \hat{a}^\dagger \hat{a}$$

$$H_{int,X}(X = A, D) = -\frac{1}{\varepsilon_0}\hat{\boldsymbol{\mu}}_{\boldsymbol{X}} \cdot \hat{\boldsymbol{d}}$$

$$\hat{\boldsymbol{d}} = i\sum_{k,\lambda}\left(\frac{\hbar c k \varepsilon_0}{2V}\right)^{1/2}\left\{\vec{\mathbf{e}}_{\boldsymbol{\lambda}}\hat{a}_{\lambda,k}e^{ikz} - \vec{\mathbf{e}}_{\lambda}^{*}\hat{a}_{\lambda,k}^{\dagger}e^{-ikz}\right\}$$

$$\hat{\boldsymbol{\mu}}_{\boldsymbol{X}}|e_X\rangle = \mu_x \vec{\mathbf{e}}_{\mathbf{x}}|g_X\rangle$$

$\hat{a}^\dagger$ and $\hat{a}$ are photonic creation and annihilation operators, and $\hat{b}_X^\dagger$ and $\hat{b}_X$ are exciton creation and annihilation operators. $\hat{\boldsymbol{d}}$ is the electric displacement operator, acting on the photonic part of eigenstates, and $\hat{\boldsymbol{\mu}}_{\boldsymbol{X}}$ is the transition dipole operator, acting on the matter part. $|e_X\rangle$ and $|g_X\rangle$ are the excited and ground state of molecule $X$, i.e., $|e_X\rangle$ represents the presence of an exciton and $|g_X\rangle$ represents the absence of an exciton.

We start by treating $H_{int}$ as a perturbative term and use second order perturbation theory, as is commonly done in RET theory[12]. The coupling strength of our system is an order of magnitude less than the exciton and cavity energies, so this approach is reasonable to get a rough approximation. We note that this light-matter interaction term is also treated perturbatively in other related theoretical works on cavity QED[10,13].

The resonant energy transfer from an initial donor state, $i$, to a final acceptor, $f$, state proceeds through two possible intermediate photonic states, $I_{1,2}$ with a total complex transition amplitude, $M$:

$$M = \frac{\langle f|H_{int}|I_1\rangle\langle I_1|H_{int}|i\rangle}{E_i - E_{I1}} + \frac{\langle f|H_{int}|I_2\rangle\langle I_2|H_{int}|i\rangle}{E_i - E_{I2}}$$

The intermediate states and their energies are listed in Table S3. $|N\rangle$ is the eigenstate of the number operator $\hat{a}^\dagger\hat{a}$ and represents a state with $N$ photons.

$M$ in total is a sum over the possible photon energies:

$$M(z_A, z_D) = \sum_{E_{ph}} M(E_{ph}, z_A, z_D)$$

*Application of Homogeneous Polaritonic Transfer Concepts*

In the limit of many photons, we replace the photon number operators with the complex electric field amplitude at a given position, $f(z)$, which can be calculated analytically with TMM. As in previous work on polaritonic transport[10], $M(E_{ph}, z_A, z_D)$ becomes the following:

$$M(E_{ph}, z_A, z_D) = \mu_A\mu_D \frac{E_{ph}}{2\varepsilon_0 V}\left\{\hat{\mathbf{e}}_{\boldsymbol{A}}^* \cdot \hat{\mathbf{e}}_{\boldsymbol{ph}} \cdot \hat{\mathbf{e}}_{\boldsymbol{D}} \cdot \hat{\mathbf{e}}_{\boldsymbol{ph}} \frac{f(z_A)f^*(z_D)}{E_D - E_{ph}} - \hat{\mathbf{e}}_{\boldsymbol{A}}^* \cdot \hat{\mathbf{e}}_{\boldsymbol{ph}}^* \cdot \hat{\mathbf{e}}_{\boldsymbol{D}} \cdot \hat{\mathbf{e}}_{\boldsymbol{ph}} \frac{f^*(z_A)f(z_D)}{E_A + E_{ph}}\right\}$$

Letting $\mu_A = \mu_D = \mu$ and $E_A = E_D = E_X + \gamma_x$ for homogenous resonance energy transfer and inclusion of exciton decay as a damping term this becomes:

$$M(E_{ph}, z_A, z_D) = \frac{\mu^2 E_{ph}}{2\varepsilon_0 V}\left\{\hat{\mathbf{e}}_{\boldsymbol{A}}^* \cdot \hat{\mathbf{e}}_{\boldsymbol{ph}} \cdot \hat{\mathbf{e}}_{\boldsymbol{D}} \cdot \hat{\mathbf{e}}_{\boldsymbol{ph}} \frac{f(z_A)f^*(z_D)}{E_X + \gamma_X - E_{ph}} - \hat{\mathbf{e}}_{\boldsymbol{A}}^* \cdot \hat{\mathbf{e}}_{\boldsymbol{ph}}^* \cdot \hat{\mathbf{e}}_{\boldsymbol{D}} \cdot \hat{\mathbf{e}}_{\boldsymbol{ph}} \frac{f^*(z_A)f(z_D)}{E_X + \gamma_X + E_{ph}}\right\}$$

The photonic decay rate, $\gamma_c$, is expected to be implicitly included through the electric field terms. We note that the light-matter coupling strength, $g$, is as follows:

$$g^2 = \frac{\mu^2\hbar\omega_c}{2\varepsilon_0 V}$$

Fitting to the coupled oscillator model suggests that $g$ is approximately invariant for the first 3 cavity modes and the interference mode (Table S1.), so we make the following substitution and approximation:

$$M(E_{ph}, z_A, z_D) \approx g^2 \left\{ \epsilon_1 \frac{f(z_A, E_{ph}) f^*(z_D, E_{ph})}{E_X + \gamma_X - E_{ph}} - \epsilon_2 \frac{f^*(z_A, E_{ph}) f(z_D, E_{ph})}{E_X + \gamma_X + E_{ph}} \right\}$$

In this expression, $\epsilon_1$ and $\epsilon_2$ are scalar constants accounting for polarization misalignment. We set both $\epsilon_1$ and $\epsilon_2$ to 1 due to the in-plane isotropy of $WS_2$ and the assumption of modes at normal incidence. The transition rate becomes an expression depending on the position of the donor:

$$\Gamma(z_D, E_{ph}) \propto \int_0^L dz_A \left| M(E_{ph}, z_A, z_D) \right|^2$$

The hopping diffusion coefficient is proportional to both the transfer rate and the length scale, and the resonance energy transfer mechanism proposed here has an inherent length scale of $|z_A - z_D|$, so we propose the following expression:

$$D(z_D, E_{ph}) \propto \int_0^L dz_A |z_A - z_D| \left| M(E_{ph}, z_A, z_D) \right|^2$$

The average diffusion coefficient is then:

$$D(E_{ph}, L) \propto \frac{1}{L} \iint_0^L dz_A dz_D |z_A - z_D| \left| M(E_{ph}, z_A, z_D) \right|^2$$

Figure S16. shows this diffusion coefficient in arbitrary units as a function of thickness and wavelength.

*Application of Theory to Fitting Quantum Efficiency Data*

The previous equation describes an average diffusion length for an exciton coupled to a photonic mode with some wavelength $\lambda_{ph}$. We assume that exciton diffusion will occur through the thickness dependent polariton modes. This is consistent with Figure S16, which shows that the diffusion term follows the polariton branches, particularly the lower polaritons. Further, this is consistent with the assumption that $M$ will be a sum over photonic modes for various wavelengths. This is expected to be the case regardless of the illumination wavelength, $\lambda_{inc}$, if $\lambda_{inc}$ can produce excitons. The efficiency of relaxation/coupling to polariton modes may differ depending on $\lambda_{inc}$. We now write the phenomenological diffusion as a sum of the contributions from each mode:

$$D(\lambda_{inc}, L) = \eta_{LP}(\lambda_{inc}) \sum_{l}^{3} D\left(E_{LP,l}(L), L\right) + \eta_{UP}(\lambda_{inc}) \sum_{l}^{3} D\left(E_{UP,l}(L), L\right)$$

$\eta_{LP,UP}(\lambda_{inc})$ represent the efficiency of lower and upper polariton creation under illumination by $\lambda_{inc}$. The Hopfield coefficients give the exciton and photon fraction of the polariton. This is effectively the probability at any given time that the polariton is in an exciton state and the fraction of the time it is in a photonic state. The polariton will only dissociate to free carriers when in the excitonic state. The diffusion coefficient is therefore scaled by the polariton exciton fraction, $|X_{LP}|^2$ for the lower polariton and $1-|X_{LP}|^2$ for the upper polariton.

Finally, we input the following expression as the

$$L_X(\lambda, L) = A_1(\lambda)\sum_{l}^{3} |X_{LP,l}|^2(L)D\big(E_{LP,l}(L), L\big) + A_2(\lambda)\sum_{l}^{3}(1-|X_{LP,l}|^2)D\big(E_{UP,l}(L), L\big)$$

The exciton diffusion current as a function of thickness and wavelength is then calculated for the resulting $L_X$, and $A_1(\lambda)$ and $A_2(\lambda)$ are the fit parameters used to fit this model to the thickness and wavelength dependent EQE in Figure 3a. Making $A_1(\lambda)$ and $A_2(\lambda)$, and therefore $L_X$, wavelength dependent accounts for the aforementioned $\eta_{LP,UP}(\lambda_{inc})$ term describing the efficiency of polariton formation.

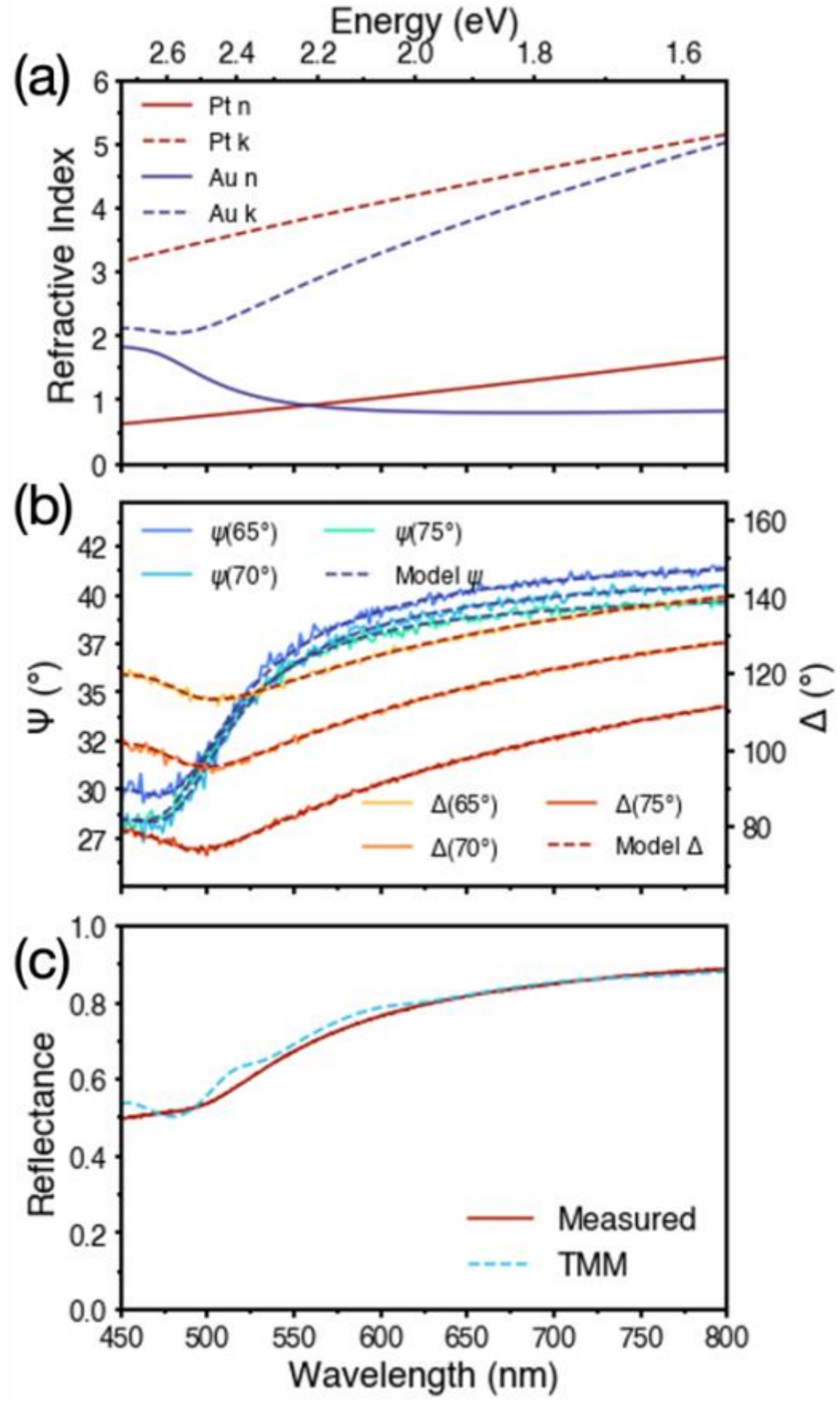

**Figure S1. Determination of Pt and Au refractive indices.** Due to the massive field enhancement from the polariton resonances, determining the refractive indices of the Pt/Au bottom electrode is crucial to effectively approximate the IQE. To get the optical constants of our Pt/Au, we performed spectroscopic ellipsometry (SE) on a Pt/Au film deposited using the exact same conditions as those used for the devices. (a) The complex refractive indices ($\tilde{n} = n + ik$) for the Pt and Au layers. (b)The measured and fit $\psi/\Delta$ values. (c) Measured and TMM[3]-calculated reflectance.

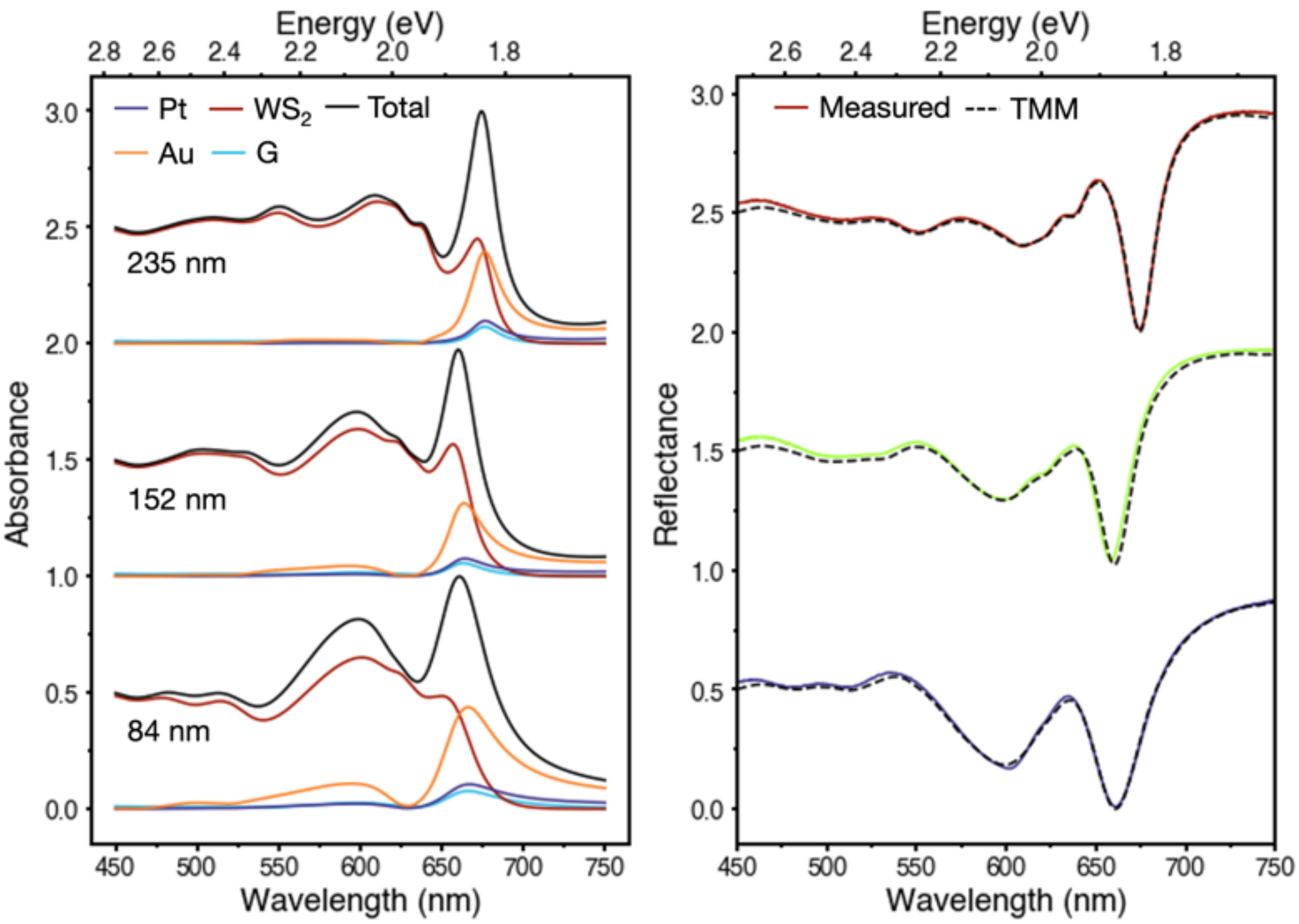

**Figure S2. Layer resolved absorption is calculated using TMM.** Examples of the calculated layer resolved absorption (left) and the experimental/calculated reflectance (right) to confirm the validity of the TMM calculations. The lossy bottom electrode layer and the field enhancement from the cavity mode result in high parasitic loss at the upper and lower polariton resonances. The parasitic loss is particularly high for the LP resonance because the $WS_2$ extinction coefficient decays below the exciton energy. The absorption at 450 nm is entirely in the $WS_2$ layer and is effectively constant with thickness, which is why it is chosen as the wavelength for off resonance measurements in Figure 3 of the main manuscript.

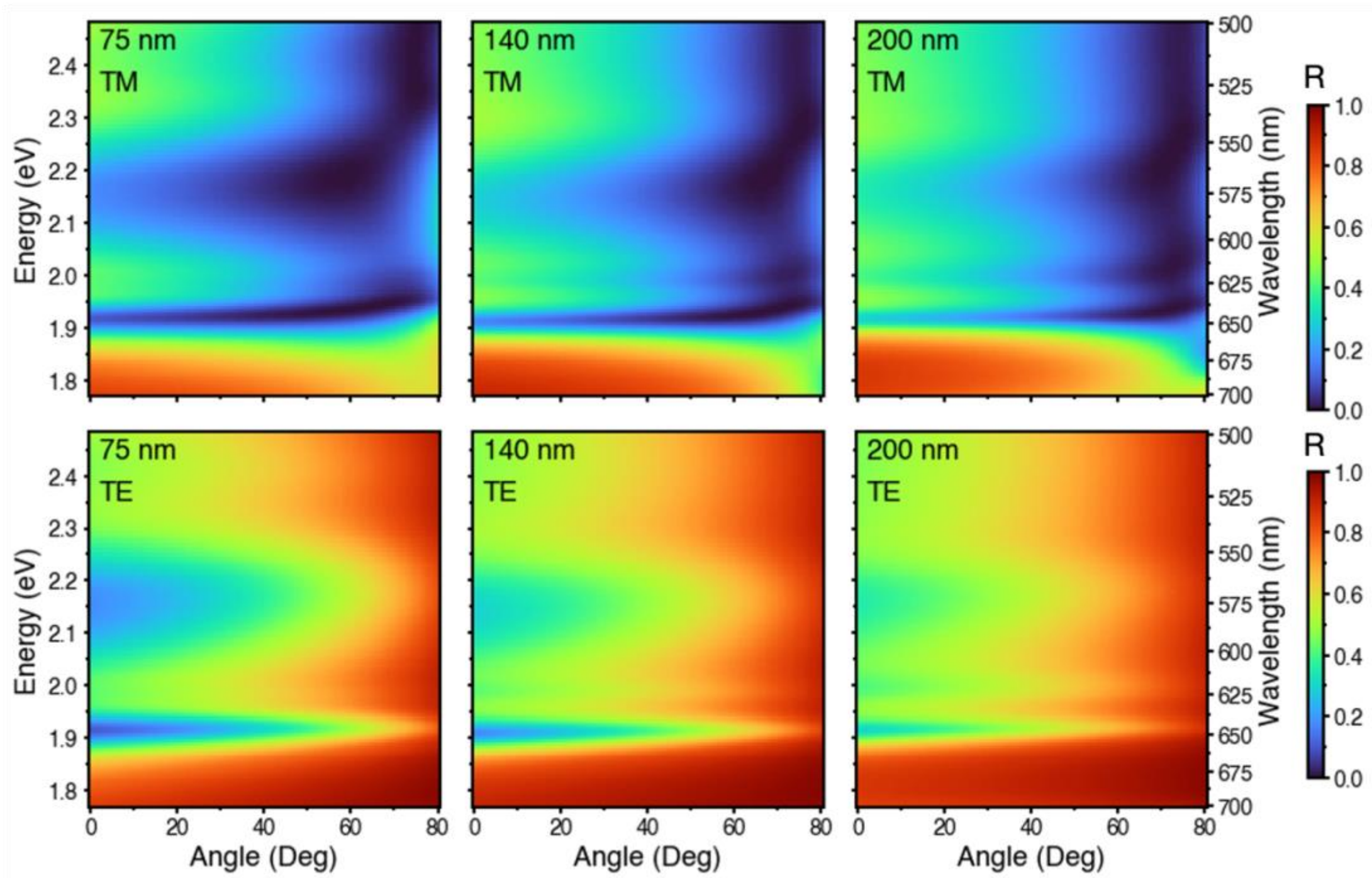

**Figure S3. Angle dependent reflectance calculations** for $WS_2$ thicknesses of 75 nm, 140 nm, and 200 nm with both TE and TM polarizations. The photon energy in a Fabry-Perot (FP) cavity depends on the path distance. Therefore, it tends to be sensitive to the incident angle. However, the path length is relatively unchanged in the self-hybridized $WS_2$ cavity system due to the large optical impedance of $WS_2$. Consequently, there is minimal dispersion for exciton-polaritons with angle, particularly for transverse electric (TE) polarizations.

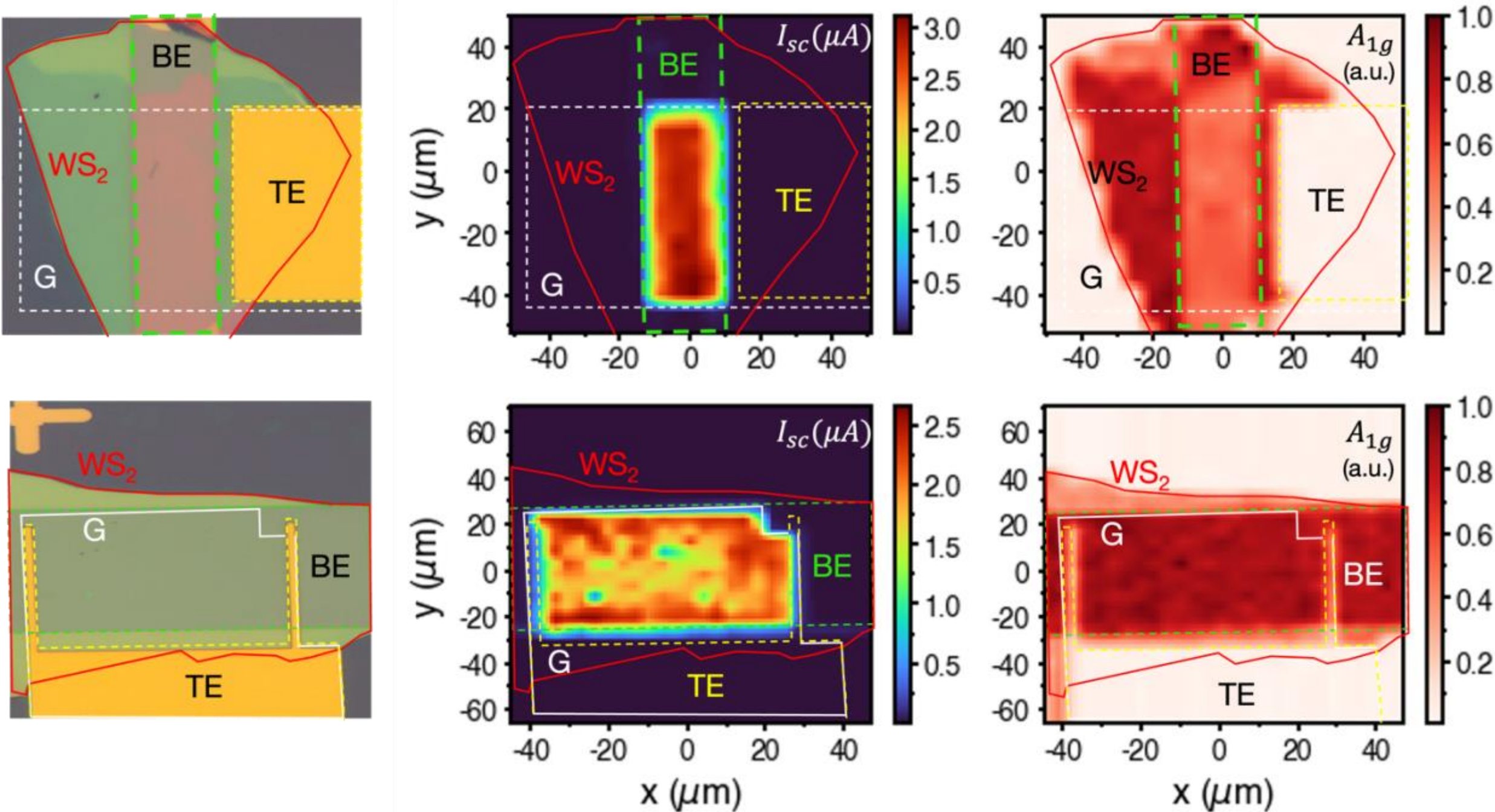

**Figure S4. Photocurrent mapping.** Optical microscope images (left), photocurrent maps (middle), and corresponding Raman maps (right) of two devices. The $WS_2$ flake is outlined in red, the bottom electrode (BE) in green, the top electrode (TE) in yellow, and the graphene (G) in white. The optical microscope images are stretched/distorted slightly to match the maps.

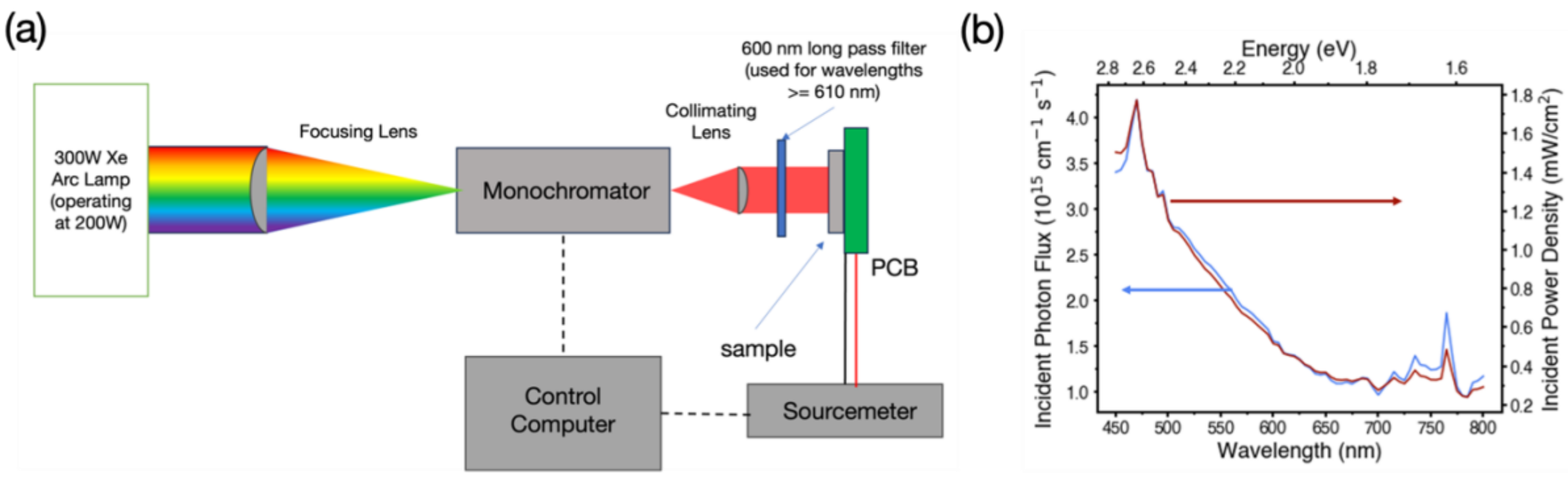

**Figure S5. Spectral EQE and PCE experimental setup.** (a) Experimental setup. (b). Baseline incident power and photon flux.

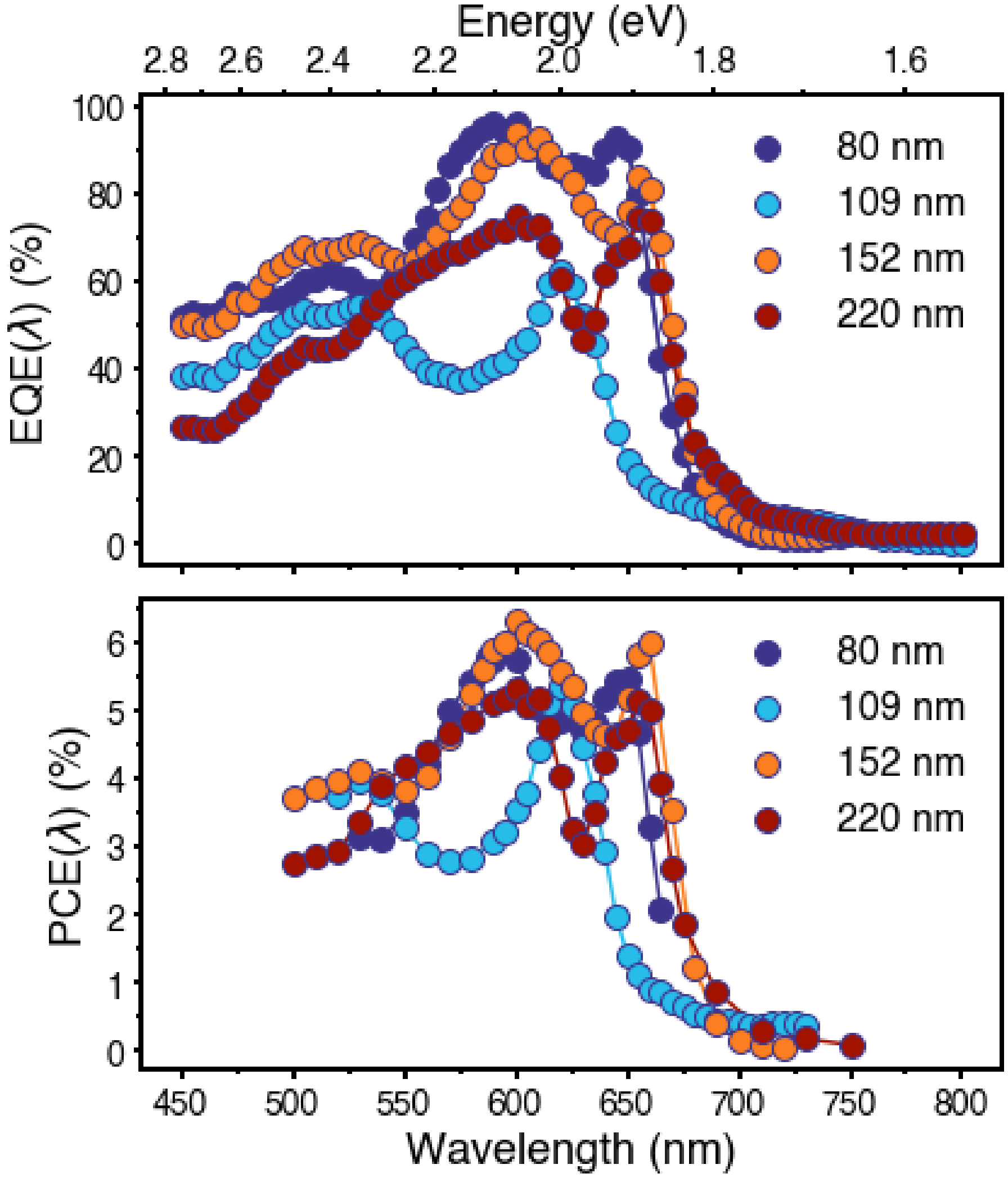

**Figure S6. Quantitative EQE and PCE spectra for representative devices.** Wavelength dependent EQE (top) and PCE (bottom) spectra for 4 representative devices with $WS_2$ thicknesses of 80,109,152, and 220 nm. We expect that the values are overestimated due to underestimation of the incident power density, which results from inhomogeneity in the power density as a function of position. We therefore choose to normalize the spectra in the main manuscript (Figure 2) and only report EQE quantitatively with a controlled spot size (Figure 3).

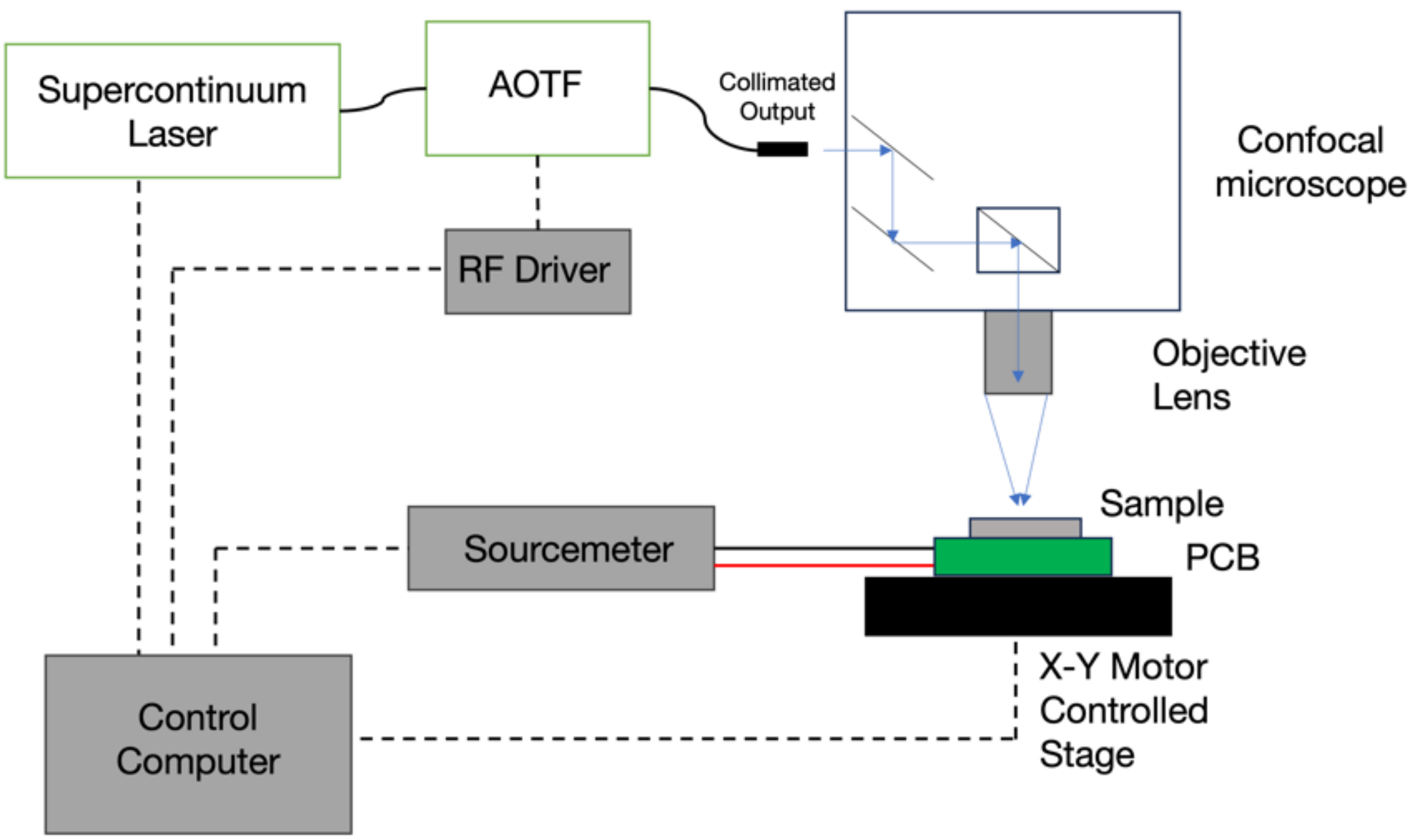

**Figure S7. Experimental setup for tunable laser EQE map measurements.** A supercontinuum laser is fiber coupled to an RF-Driven acousto-optic tunable filter, the output of which is fiber coupled. The fiber output is collimated and enters a confocal microscope where a beam splitter couples the light into a 50x 0.35 NA lens. The mapping is done using an automated stage.

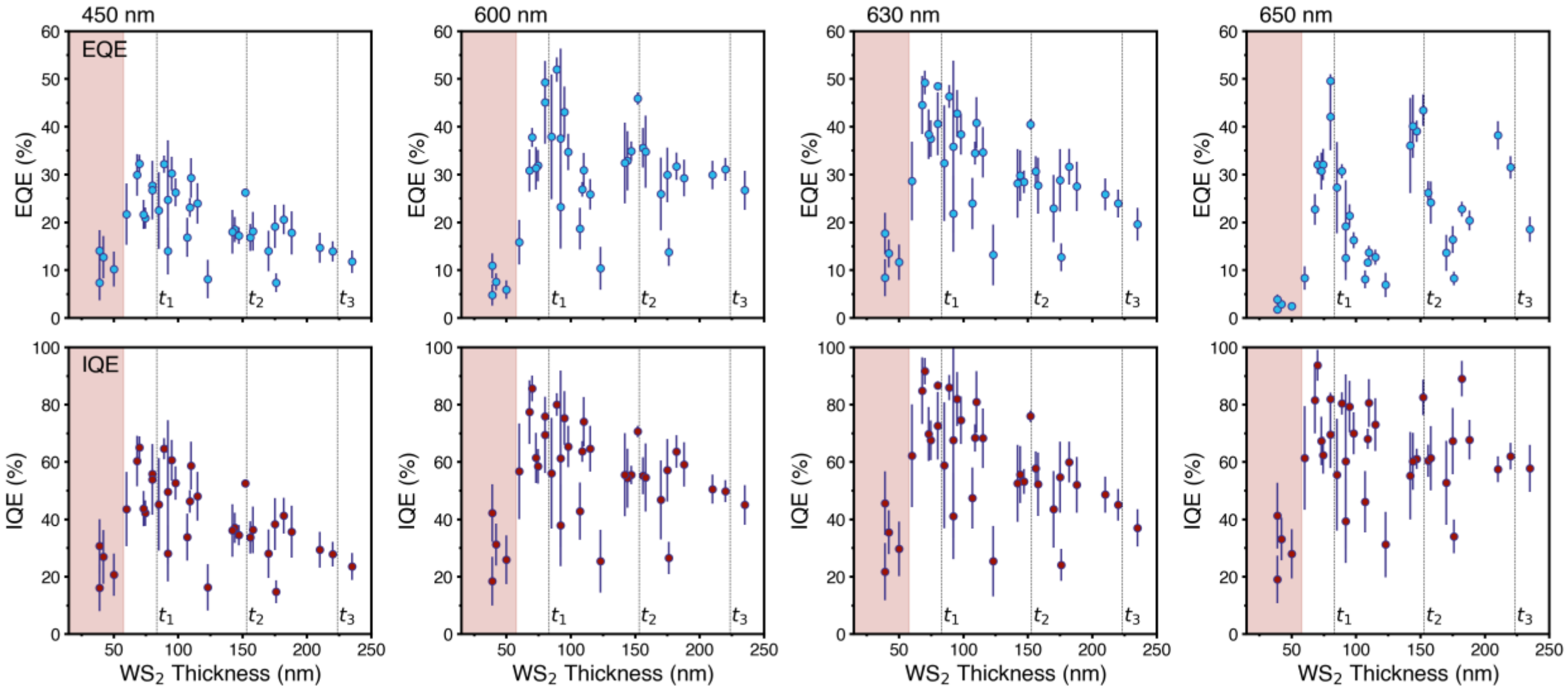

**Figure S8. All thickness dependent EQE and IQE data points**

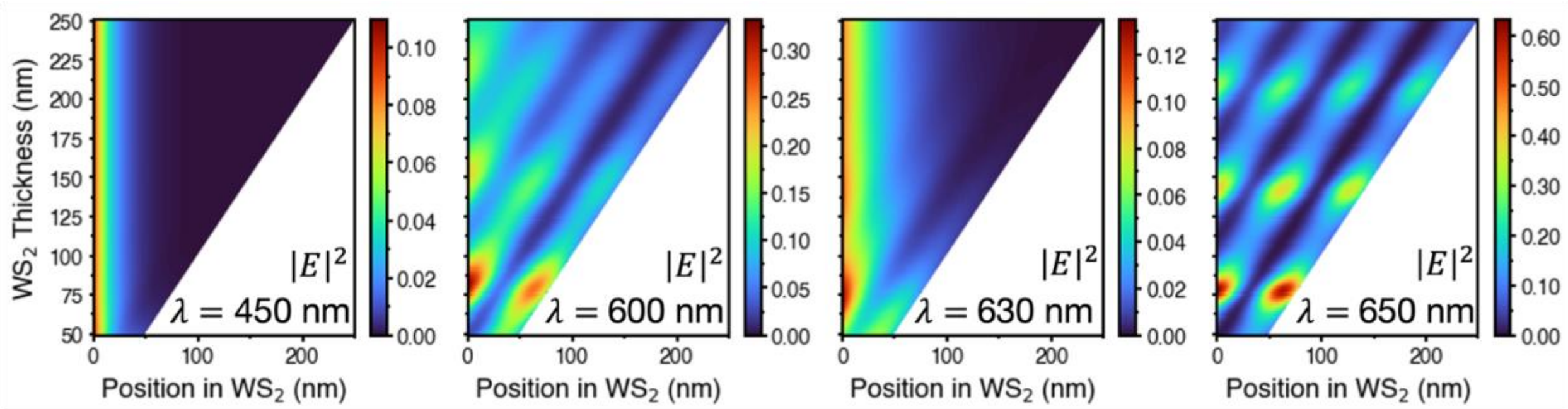

**Figure S9. Thickness and position dependent electric field.** Position dependent electric field magnitude as a function of $WS_2$ thickness as calculated by TMM for incident wavelengths of 450 nm, 600 nm, 630 nm, and 650 nm. The position dependent electric field magnitude, $|\boldsymbol{E}|^2$, is shown as a function of thickness for the four wavelengths of interest. $|\boldsymbol{E}|^2$ is directly proportional to power absorption and therefore generation rate. Near polariton resonances in the plots for 600 nm and 650 nm, the electric field magnitude is increased near the z = 0 and z = L of the $WS_2$ (i.e., near each electrode). The electric field for the 450 nm and 630 nm cases are effectively invariant with thickness.

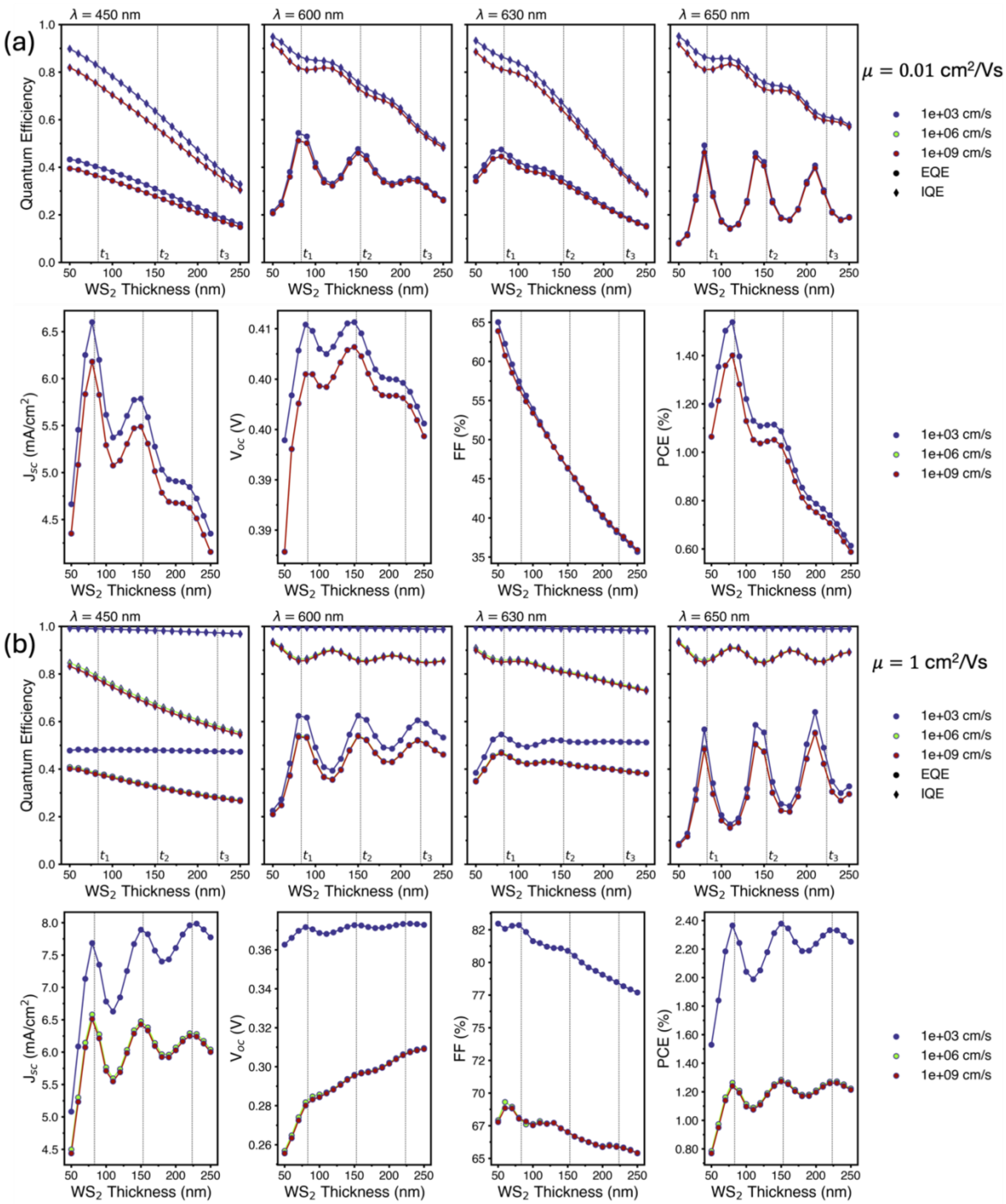

**Figure S10. Effect of surface recombination velocity** on thickness dependent EQE/IQE and PV parameters for mobilities of (a) 0.01 $cm^2$/Vs, (b) 1 $cm^2$/Vs. The carrier lifetimes are constant at 50 ns and the front/back work functions are 4.4 eV and 4.9 eV. The effects of surface recombination appear to saturate at/beyond SRVs of $10^6$ cm/s. Surface recombination does not explain the increase in IQE corresponding to the $t_1$ peak in Figure 3, nor does it explain the increase in FF in the strong coupling regime in Figure 4e.

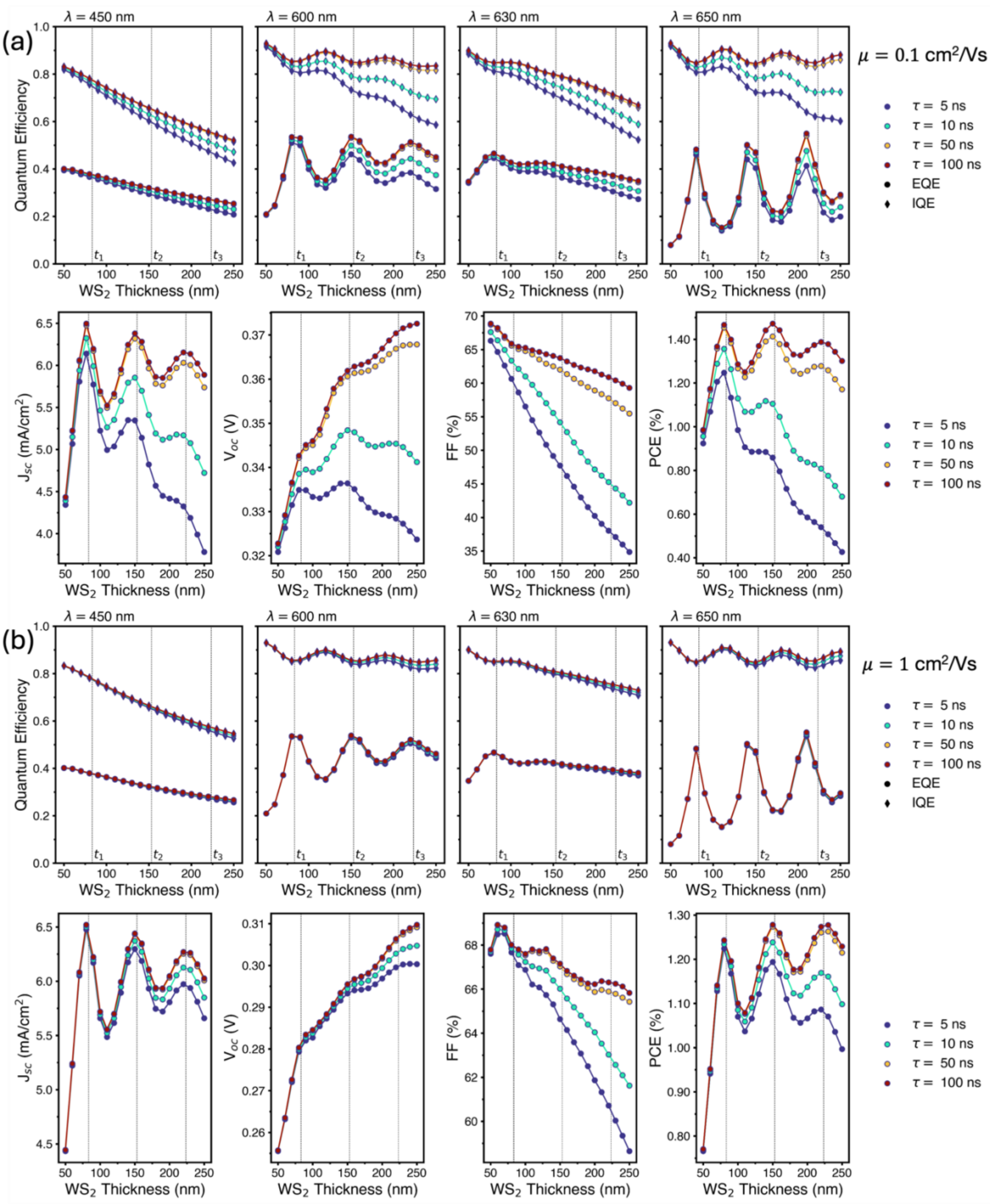

**Figure S11. Effect of SRH limited lifetime** on EQE/IQE and PV parameters for various carrier mobility values of (a) 0.01 cm$^2$/Vs, (b) 1 cm$^2$/Vs. The carrier lifetimes are constant at 50 ns and the front/back work functions are 4.4 eV and 4.9 eV.

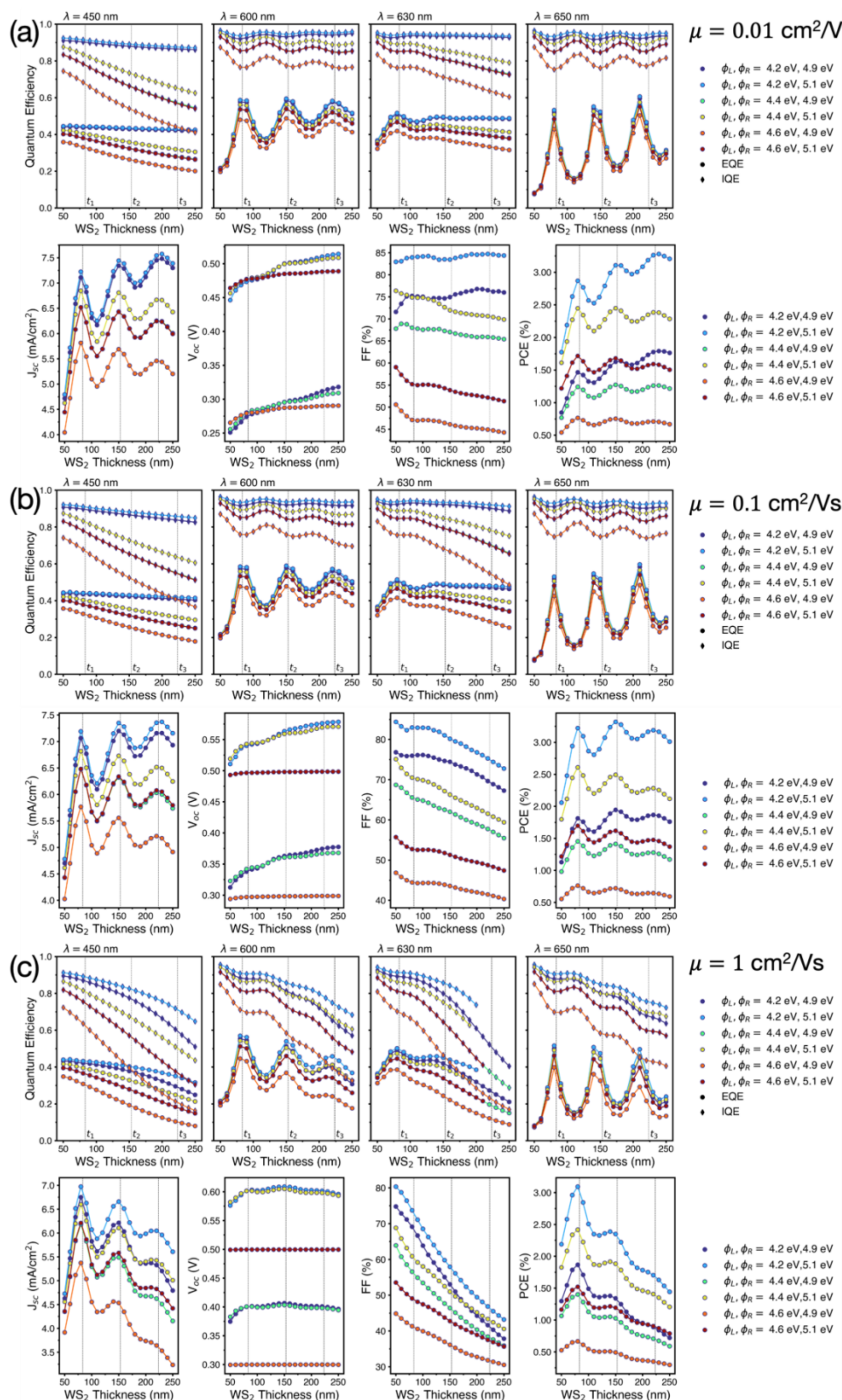

**Figure S12. Effect of front and back work function** on thickness dependent EQE/IQE and PV parameters for carrier lifetimes of 50 ns and mobilities of (a) 0.01 cm$^2$/Vs, (b) 0.1 cm$^2$/Vs, and (c) 1 cm$^2$/Vs.

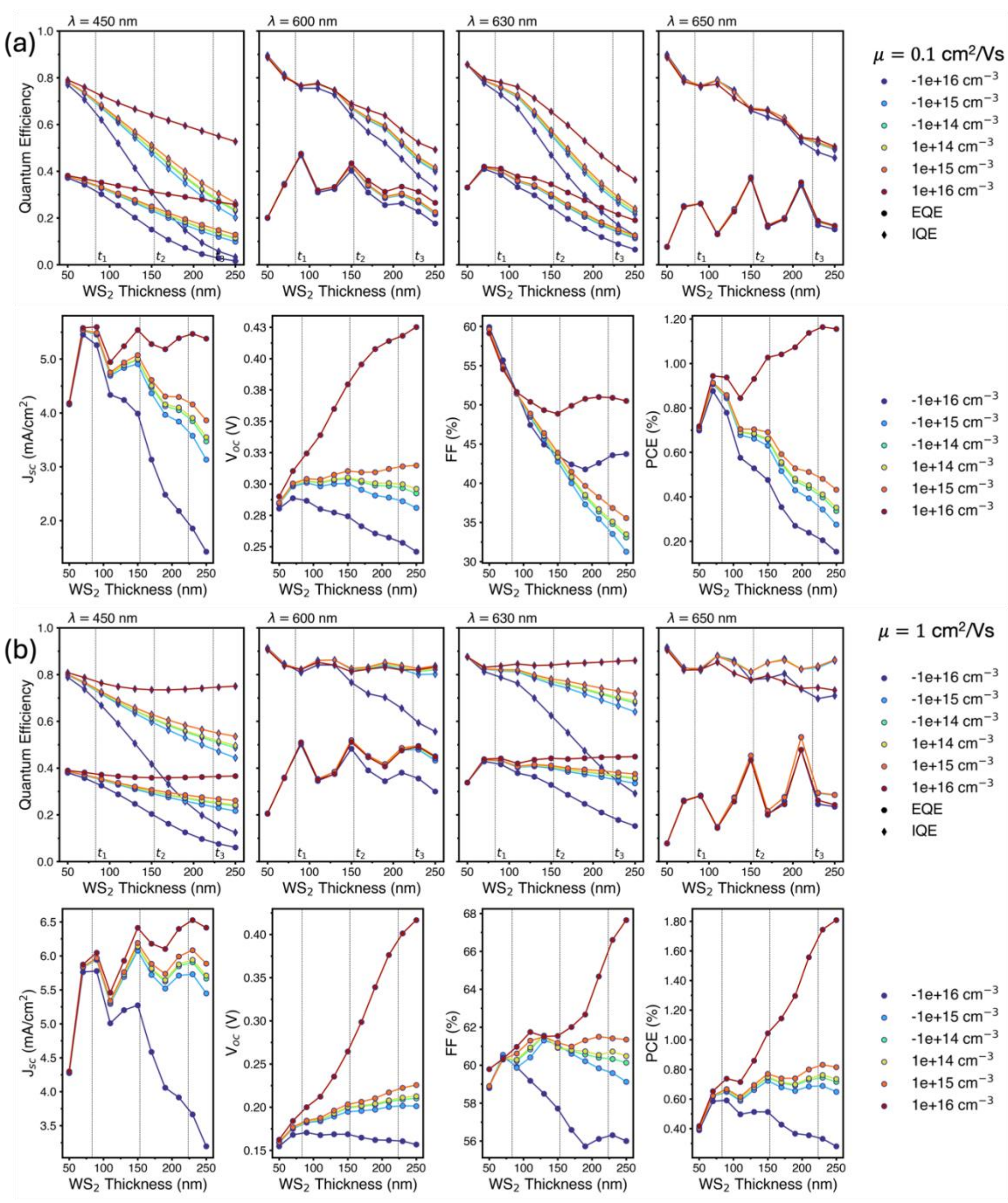

**Figure S13. Effect of doping** on thickness dependent EQE/IQE and PV parameters for carrier lifetimes of 50 ns and mobilities of (a) 0.1 cm$^2$/Vs, and (b) 1 cm$^2$/Vs. Negative values indicate n-doping and positive values indicate p-doping.

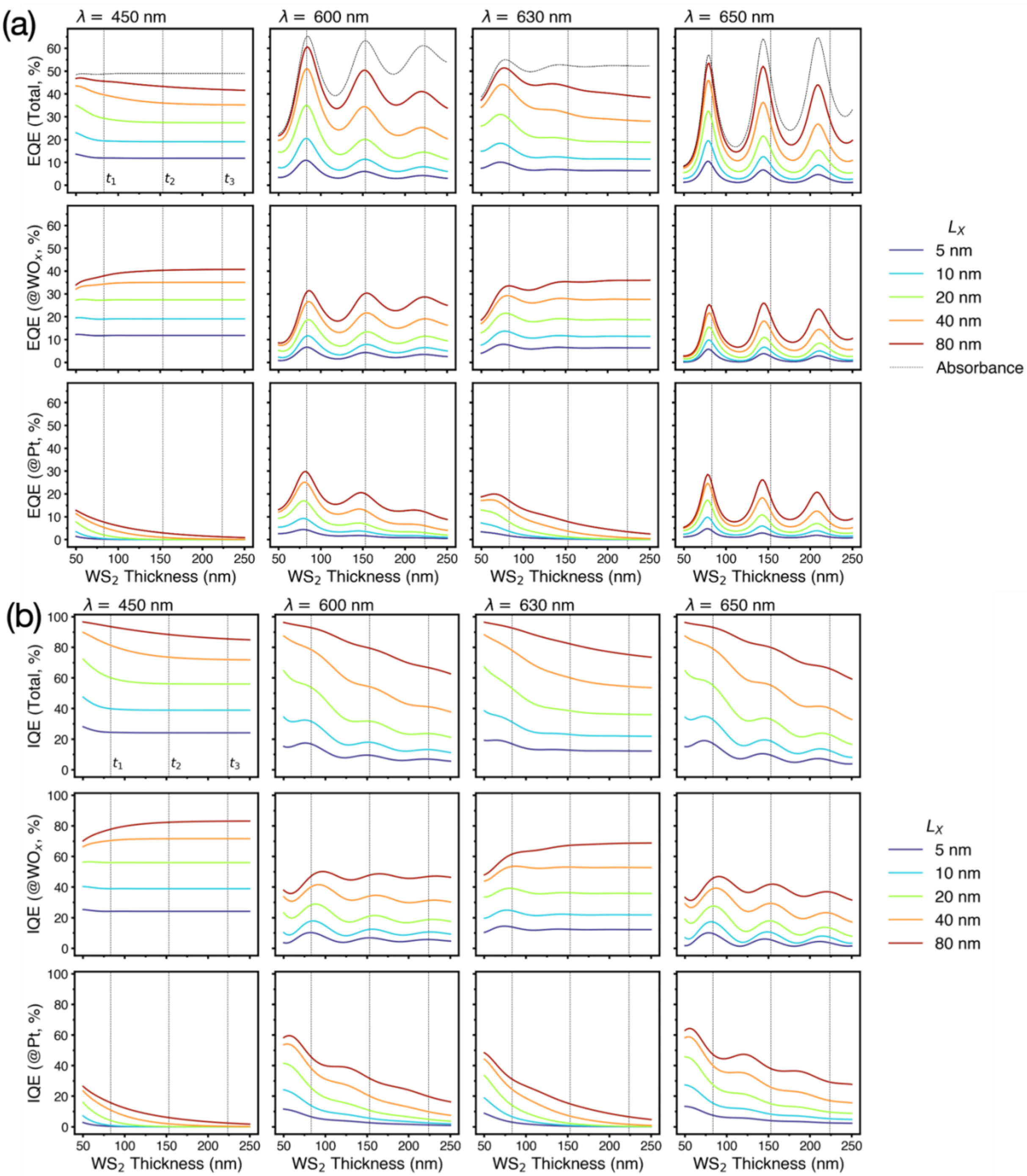

**Figure S14. Excitonic photocurrent.** Thickness dependence of (a) EQE and (b) IQE resulting from excitonic photocurrent under illumination by (left to right) 450,600,630, and 650 nm light for various exciton diffusion lengths. For both EQE and IQE, the total QE assuming perfect collection at each interface is plotted in the first row, then the QE resulting exclusively from collection at the front ($WO_x$) and back (Pt) interfaces are considered to account for possible interfacial effects. This model does not replicate our results in Figure 3 of the main manuscript for any combination of interfacial dissociation efficiencies or thickness independent exciton diffusion length.

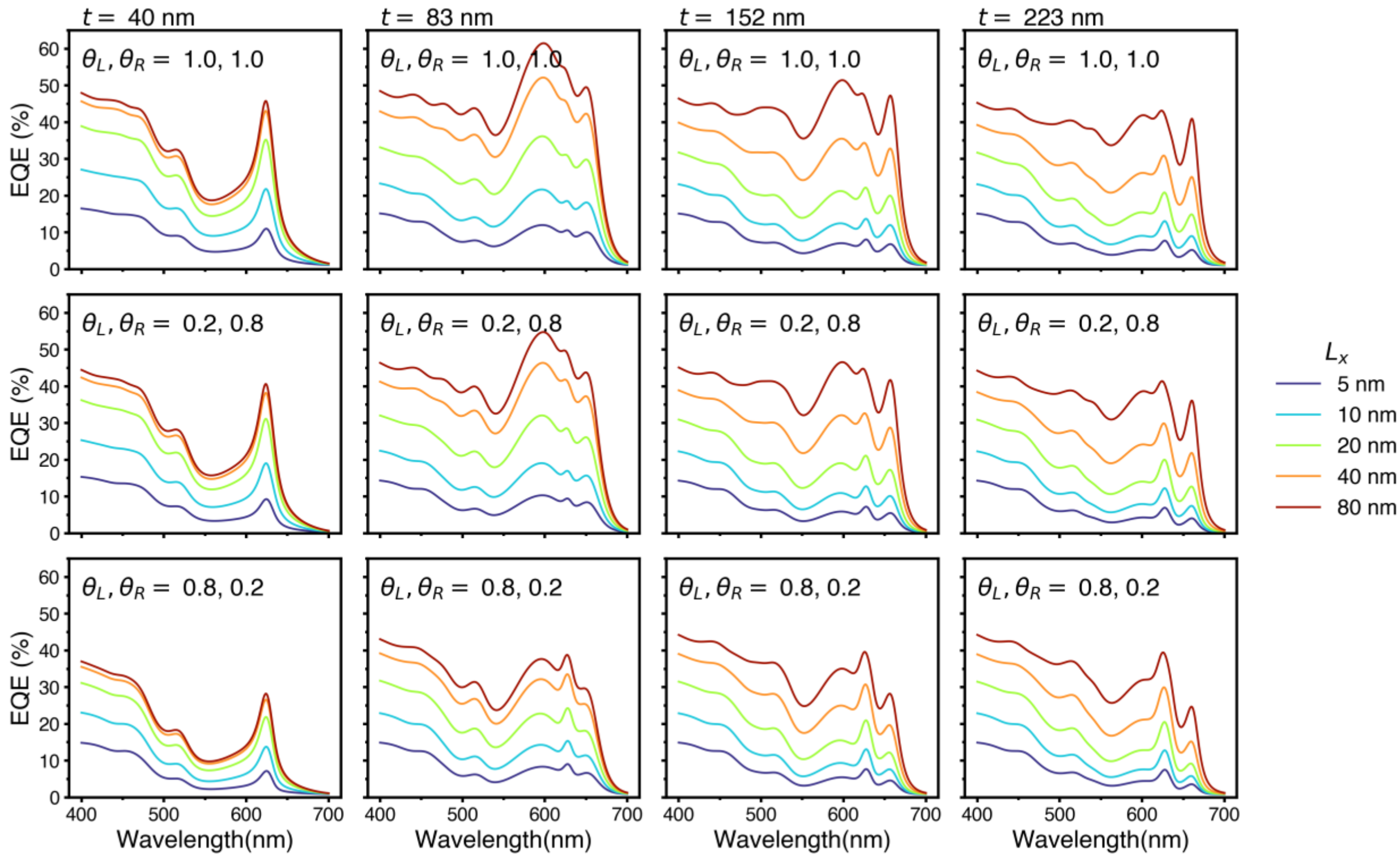

**Figure S15. Excitonic photocurrent EQE spectra** as a function of exciton diffusion length calculated for thicknesses of (left to right) 40, 83, 152, and 223 nm thickness and three pairs of interfacial exciton dissociation quantum efficiencies (top to bottom). The results differ from the experimental results in Figure 2a of the main manuscript.

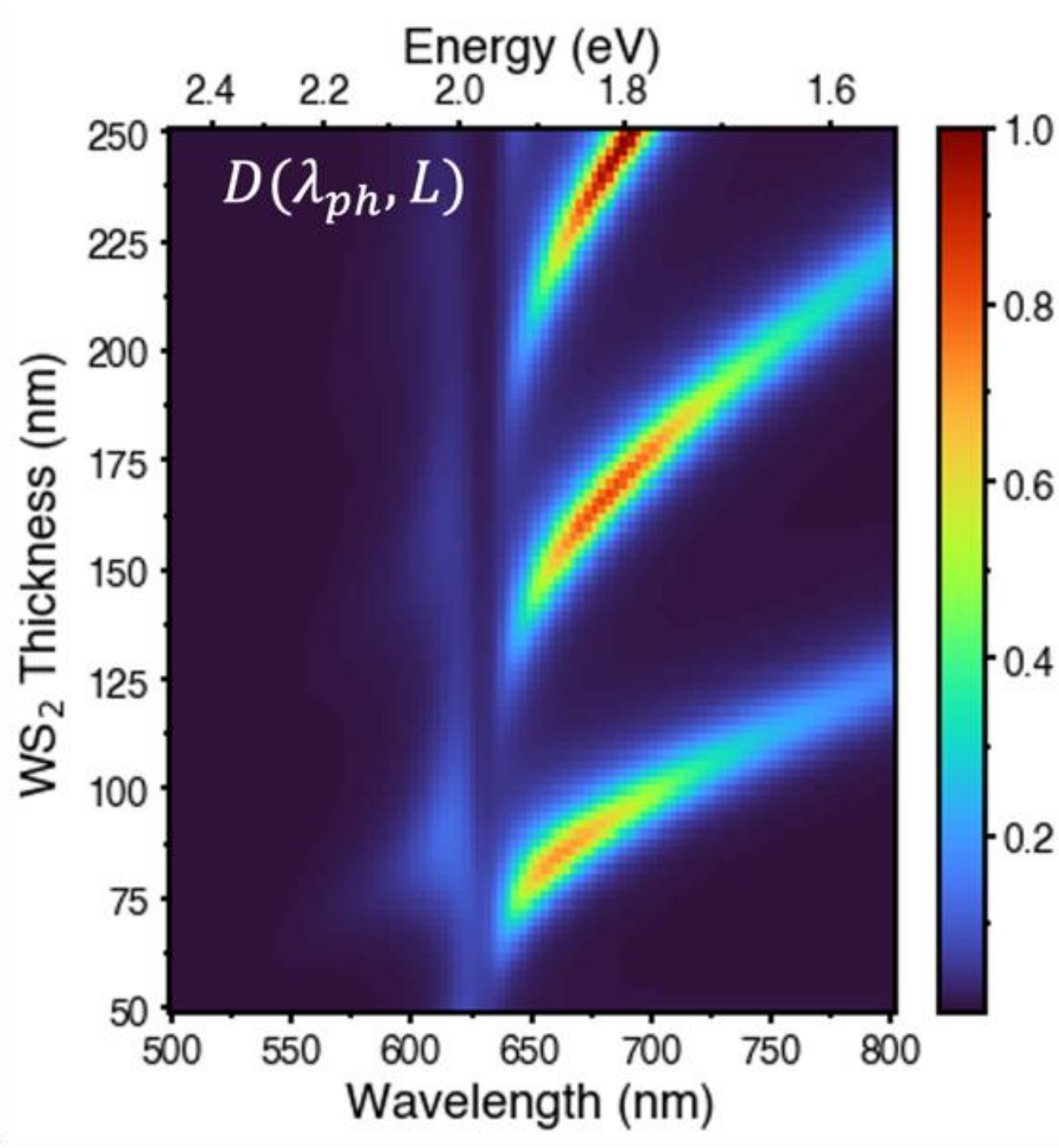

**Figure S16. Phenomenological diffusion coefficient in arbitrary units.**

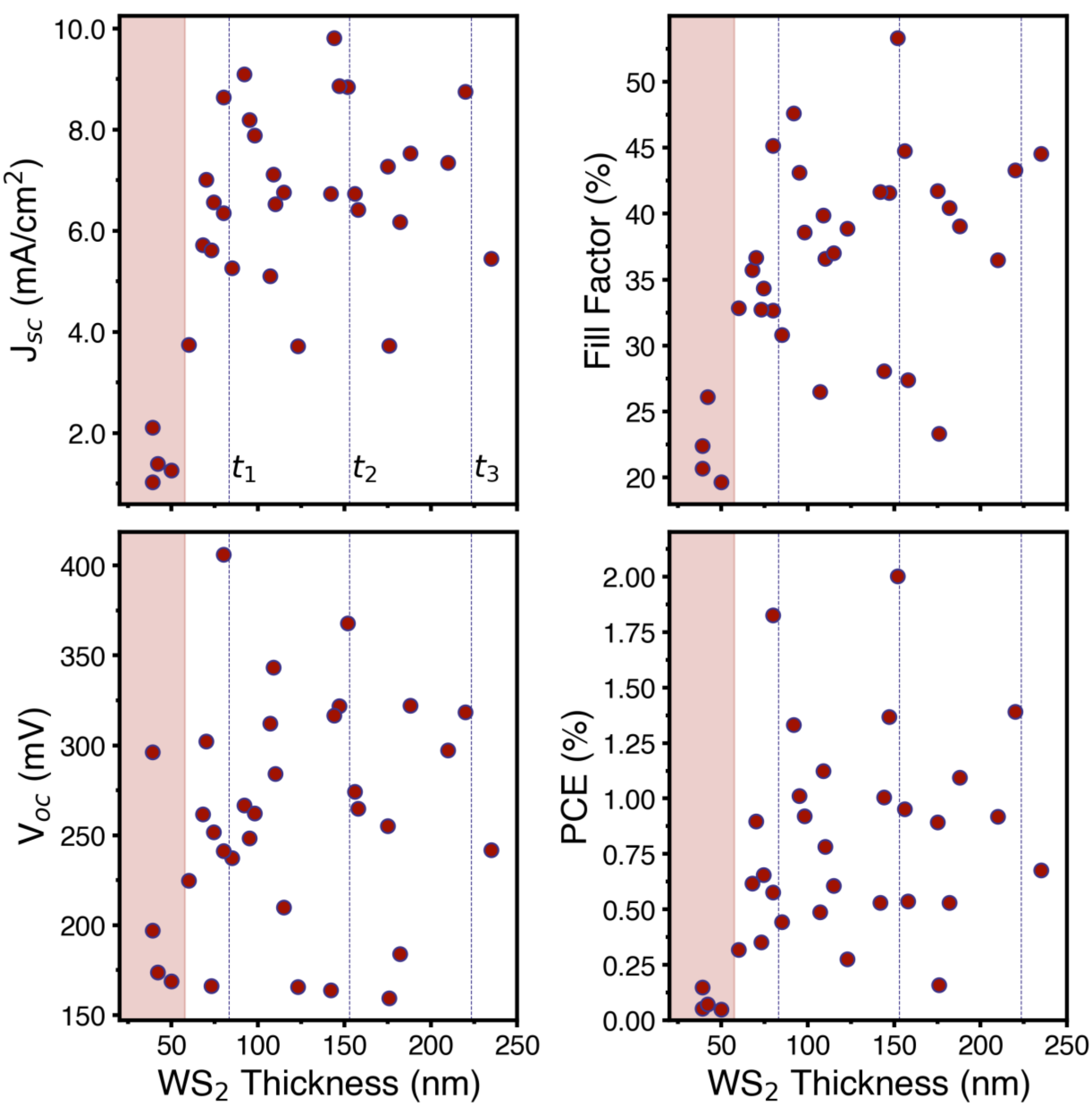

**Figure S17. PV characteristics of all devices.**

**Table S1. Coupled oscillator model fit results for *l* = 0,1,2,3,4 cavity mode orders.** The Rabi splitting and the coupling regime for the cavity mode coupling with the A exciton are given.

| Parameter | Mode Order | | | | |
|---|---|---|---|---|---|
| | $l = 0$ | $l = 1$ | $l = 2$ | $l = 3$ | $l = 4$ |
| $g_A$ (meV) | 99.8 | 113.8 | 104.3 | 100.4 | 97.9 |
| $g_{lB}$ (meV) | 78.4 | 77.7 | 78.2 | 87.0 | 89.3 |
| $g_{lC}$ (meV) | 49.136 | 49.1 | 49.4 | - | - |
| $E_X^{(A)}$ (eV) | 1.966 | 1.968 | 1.969 | 1.968 | 1.967 |
| $E_X^{(B)}$ (eV) | 2.386 | 2.385 | 2.385 | 2.385 | 2.385 |
| $E_X^{(C)}$ (eV) | 2.690 | 2.690 | 2.690 | - | - |
| Cavity Slope | 12.500 | 4.346 | 2.562 | 1.812 | 1.426 |
| Cavity Intercept (nm) | 462.5 | 260.9 | 233.3 | 221.8 | 207.1 |
| $\gamma_C$ (meV) | 327.7 | 145.9 | 95.3 | 71.0 | 55.1 |
| $\gamma_X^{(A)}$ (meV) | 25.1 | 30.0 | 26.1 | 27.4 | 25.0 |
| $\gamma_X^{(B)}$ (meV) | 100.0 | 99.8 | 100.0 | 102.0 | 100.0 |
| $\gamma_X^{(C)}$ (meV) | 125.0 | 124.8 | 125.0 | - | - |
| **Rabi Splitting at $X_A$ (meV)** | **-** | **195.8** | **196.9** | **196.0** | **193.5** |
| **Coupling Regime** | Weak | Strong | Strong | Strong | Strong |

**Table S2. Power density of light incident on sample vs wavelength for tunable laser EQE map measurements.**

| Wavelength (nm) | Optical Power Density ($W/cm^2$) |
|---|---|
| 450 | 2.29 |
| 600 | 5.09 |
| 630 | 5.12 |
| 650 | 5.86 |

**Table S3. Possible intermediate states in the RET framework and their respective energies.**

| State | Energy |
|---|---|
| $\lvert i \rangle = \lvert e_D, g_A, N \rangle$ | $E_D$ |
| $\lvert I_1 \rangle = \lvert g_D, g_A, N+1 \rangle$ | $E_{ph}$ |
| $\lvert I_2 \rangle = \lvert e_D, e_A, N+1 \rangle$ | $E_D + E_A + E_{ph}$ |
| $\lvert f \rangle = \lvert g_D, e_A, N \rangle$ | $E_A$ |

**Table S4a. Drift-Diffusion Simulation Materials Parameters.**[5,14,15] In some cases, ranges of values are given for parameter sweeps.

| | $WS_2$ |
|---|---|
| Thickness (nm) | 50-250 |
| Electron Affinity $\chi$ (eV) | 4.2 |
| Band gap $E_g$ (eV) | 1.36 |
| $\varepsilon_r$ | 5.8 |
| $\mu_e$ ($cm^2/Vs$) | $10^{-2}$-$10^{0}$ |
| $\mu_h$ ($cm^2/Vs$) | $10^{-2}$-$10^{0}$ |
| $N_C$ ($cm^{-3}$) | $1.26\times10^{19}$ |
| $N_V$ ($cm^{-3}$) | $1.93\times10^{19}$ |
| $\tau_{SRH}$ (s) | $10^{-8}$-$10^{-7}$ |